\documentclass[superscriptaddress,amsmath,amssymb, aps, prx,longbibliography,nofootinbib,twocolumn]{revtex4-2}
\usepackage[normalem]{ulem}
\usepackage{float}
\usepackage{bm}
\usepackage{xcolor}
\usepackage[dvipsnames]{xcolor}
\usepackage{mathtools}  
\usepackage{graphicx}
\usepackage{hyperref}

\hypersetup{colorlinks,breaklinks,
            urlcolor=[rgb]{0.08,0.03,1},
            linkcolor=[rgb]{0.08,0.03,1},
            citecolor=[rgb]{0.08,0.03,1}}

\usepackage{booktabs} % better tables

\newcommand{\fsl}[1]{\ensuremath{\mathrlap{\not{\phantom{#1}}}#1}}

\makeatletter
\renewcommand\footnotesize{%
   \@setfontsize\footnotesize\@ixpt{11}%
   \abovedisplayskip 1\p@ \@plus2\p@ \@minus4\p@
   \abovedisplayshortskip \z@ \@plus\p@
   \belowdisplayshortskip 2\p@ \@plus2\p@ \@minus2\p@
   \def\@listi{\leftmargin\leftmargini
               \topsep 2\p@ \@plus2\p@ \@minus2\p@
               \parsep 1\p@ \@plus\p@ \@minus\p@
               \itemsep \parsep}%
   \belowdisplayskip \abovedisplayskip
}
\makeatother

\newcommand{\Harvard}{Department of Physics, Harvard University, Cambridge, Massachusetts 02138, USA.}
\newcommand{\TUM}{Technical University of Munich, TUM School of Natural Sciences, Physics Department, 85748 Garching, Germany}
\newcommand{\MCQST}{Munich Center for Quantum Science and Technology (MCQST), Schellingstr. 4, 80799 M{\"u}nchen, Germany}

\begin{document}

\title{Microscopic Mechanism of Anyon Superconductivity Emerging from Fractional Chern Insulators}

\author{Fabian Pichler}
\thanks{These authors contributed equally to this work.\\}
\affiliation{\TUM}
\affiliation{\MCQST}
\author{Clemens Kuhlenkamp}
\thanks{These authors contributed equally to this work.\\}
\affiliation{\Harvard}
\author{Michael Knap}
\affiliation{\TUM}
\affiliation{\MCQST}
\author{Ashvin Vishwanath}
\affiliation{\Harvard}

\date{\today}

\begin{abstract}
Fractional quantum Hall (FQH) states and superconductors typically require contrasting conditions, yet recent experiments have observed them in the same device. A natural explanation is that mobile anyons give rise to superconductivity; however, this mechanism requires binding of minimally charged anyons to establish an unusual energy hierarchy. This scenario has mostly been studied with effective theories, leaving open the question of how anyon superconductivity can arise from repulsive interactions. Here, we show that such an energy hierarchy of anyons arises naturally in fractional Chern insulators (FCIs) at fillings $\nu = 2/(4p \mp 1)$ when they are driven toward a quantum phase transition into a ``semion crystal''---an exotic charge-density-wave (CDW) insulator with semion topological order. Near the transition, Cooper-pair correlations are enhanced, so that a conventional charge-2e superconductor appears with doping. Guided by these insights, we analyze a microscopic realization in a repulsive Hubbard-Hofstadter model. Tensor network simulations at $\nu = 2/3$ reveal a robust FCI that, with increasing interactions, transitions into the semion crystal. Finding a stable semion crystal in such a minimal model  highlights it as a viable state competing with conventional CDW and FQH states. In the vicinity of this transition, we find markedly enhanced Cooper pairing, consistent with our theory that the 2e/3 anyon is cheaper than a pair of isolated e/3 anyons. Doping near the transition should in general lead to doping Cooper pairs and charge-2e superconductivity, with chiral edge modes of alternating central charge $c = \pm2$, which can coexist with translation symmetry breaking. Our framework unifies recent approaches to anyon superconductivity, reconciles it with strong repulsion and provides guidance for flat band moir\'e materials such as recent experiments in twisted MoTe$_2$.

\end{abstract}

\maketitle

\tableofcontents

\section{Introduction}

\begin{figure*}
\begin{center}
\includegraphics[width=0.99\linewidth]{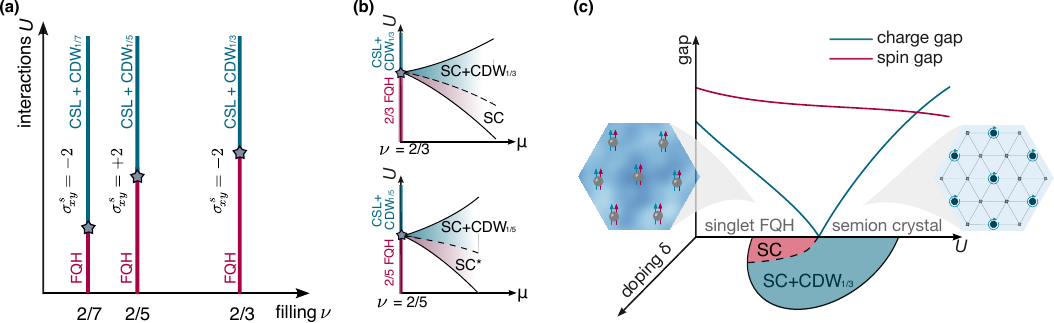}
\caption{\textbf{Setup and phase diagram.} \textbf{(a)} 
At various fractional fillings seemingly continuous topological phase transitions arise between FQH states and semion crystals---charge density wave states which support a chiral spin liquid. The sign of the spin Hall conductivity alternates with the filling of the FQH state. \textbf{(b)} Upon changing the chemical potential, a conventional superconducting (SC) state of anyons with chiral edge states arises that potentially coexists with density wave order. Two cases of 2/3 (top) and 2/5 (bottom) filling are illustrated. \textbf{(c)} At the transition from a singlet FQH state to the semion crystal, the charge gap closes, while the spin gap remains finite across the transition. Consequently, it is favorable for doped charges to enter as charge-2 singlets {of zero spin}, providing a microscopic mechanism for superconductivity upon doping close to the critical point.}
\label{fig:overview}
\end{center}
\end{figure*}

\label{sec:Introduction}

Anyons, quasiparticles with fractional statistics, and often with fractional charge, are the emergent low-energy excitations of strongly correlated systems with topological order~\cite{wen_book_2007, simon2023topological}. 
The first experimental realization of anyons occurred in fractional quantum Hall (FQH) states, discovered in fractionally filled Landau levels~\cite{Tsui1982}. However,  charged anyons in the Landau levels of conventional FQH states lack kinetic energy, due to the strong magnetic field. This enables the formation of incompressible FQH plateaus, but restricts dynamics and prohibits the emergence of competing phases, such as superconductors~\cite{Spodyneiko2023}. Yet, a dispersion can be imprinted on anyons when a periodic potential is applied to a Landau level~\cite{KolRead1993,Tang_SC_2013} or when the fractional state results from a Chern band{~\cite{Sorensen2005,Hafezi2007,Sheng_FractionalQuantum_2011, Neupert_FractionalQuantum_2011, regnault_FractionalChern_2011}}. It was realized early on that a gas of charged anyons with a finite dispersion may form a superconductor~\cite{laughlinRelationshipHighTemperatureSuperconductivity1988, laughlinSuperconductingGroundState1988, fetterRandomphaseApproximationFractionalstatistics1989, leeAnyonSuperconductivityFractional1989, wenChiralSpinStates1989, hosotaniSuperconductivityAnyonModel1990, wenCompressibilitySuperfluidityFractionalstatistics1990, leeANYONSUPERCONDUCTIVITYCHARGEVORTEX2012,chenAnyonSuperconductivity1989}. Recent experiments with two-dimensional semiconductor heterostructures have indeed observed both fractional Chern insulators stabilized by magnetic fields~\cite{Spanton2018, Yacobi2021} and Fractional Quantum Anomalous Hall (FQAH) states at zero fields~\cite{cai_SignaturesFractional_2023, zeng_ThermodynamicEvidence_2023,park_ObservationFractionally_2023, xu_ObservationInteger_2023,LongJu2024}. FQAH states are thus realized on a lattice and are often stable down to zero magnetic field. These properties endow the anyons with finite kinetic energy, opening up the possibility for the formation of competing states, not available in continuum Landau levels~\cite{Tang_SC_2013, Lee2018, schoonderwoerd2022, Shi2024, kim2025}. Several recent developments in moir\'e materials~\cite{LongJuSc2025}, including the observation of superconductivity in MoTe${_2}$~\cite{xu_SC_2025}, the same material in which FQAH states have been realized~\cite{cai_SignaturesFractional_2023, zeng_ThermodynamicEvidence_2023,park_ObservationFractionally_2023, xu_ObservationInteger_2023}, has rekindled interest in connections between  superconductivity and anyons~\cite{kim2025, Shi2024, divic_anyonsuperconductivity_2024}.

A key ingredient in determining whether a liquid of charged anyons {\em actually} leads to a superconductor hinges on the choice of anyon. Recent theoretical studies of doping anyons point to the following simple diagnostic, at least for abelian anyons. Fuse the anyon repeatedly until a local excitation (a finite group of electrons) appears. If the first local excitation is bosonic (e.g. a Cooper pair), a doped gas is expected to superconduct; if instead the first local excitation is fermionic (e.g., the electron), superconductivity is pre-empted by other phases. Indeed the charge-e semion of Laughlin's original proposal~\cite{laughlinRelationshipHighTemperatureSuperconductivity1988,laughlinSuperconductingGroundState1988} and in~\cite{divic_anyonsuperconductivity_2024}, as well the~\cite{Shi2024} charge-2e/3 anyon of the Laughlin $\nu=1/3$ state satisfy this condition by fusing with themselves to directly give a charge-2e boson; a finite density of these excitations is an ``anyon superconductor'', in contrast to doping the charge-e/3 anyon of the Laughlin state. However, the latter are generally believed to be energetically cheaper given the predominantly repulsive interactions and their minimal charge, and are more commonly realized as dopants in experiments. Hence, the central challenge in engineering superconductivity out of fractional Chern insulators (FCIs) is energetics: one must arrange for the right anyon to be the cheapest charge carrier, and the one activated on doping. Consequently, in the previous example, we must ensure that the energy cost of a 2e/3 anyon is lower than that of two well-separated e/3 anyons. In this paper, we provide a general mechanism for realizing this energetic hierarchy. 

Simply put, our mechanism invokes a third phase, in addition to the FCI and superconductor. This phase is an insulator but with semion topological order, which for general filling breaks translation symmetry---hence we call it a semion crystal (SX). We will argue that if the FCI is in proximity to the SX, more precisely if it is near a quantum critical point into that phase, then one is guaranteed to have the correct anyon energetics to favor superconductivity. Such a topological criticality mechanism for superconductivity has recently been discussed for {\em integer} Chern states adjacent to chiral spin liquids~\cite{divic_anyonsuperconductivity_2024}, which provides a modern perspective on anyon superconductivity of charge-e semions~\cite{laughlinSuperconductingGroundState1988,laughlinRelationshipHighTemperatureSuperconductivity1988}.
 The semion crystal phase is quite natural in the FCI context. It has previously come up to explain insulators with vanishing Chern response observed in MoTe$_2$~\cite{Song2024}. {The appearance of a CDW with semion topological order seems surprising at first sight; the following viewpoint may help clarify why it emerges here. In the same triangular lattice Hubbard model with $\pi/2$ flux per plaquette, at unit filling per site, a chiral spin liquid (CSL) is well established \cite{Kuhlenkamp24}. For the case considered in this work, at 2/3 filling the system instead breaks translation symmetry forming a new triangular lattice with a three times larger unit cell, such that each elementary plaquette now encloses $3\pi/2=-\pi/2 {\,\rm (mod} \, 2\pi \rm) $ flux. The effective lattice again favors a CSL, but with reversed chirality, thus rationalizing both the emergence of the SX and the predicted opposite  chirality of the topological order.}
 
 While not essential, these discussions are most transparent when the FQH states arise from electrons with a two-component spin index, which may correspond to physical spin or an approximate pseudo-spin, such as a layer or valley index~\cite{ZhangSU4_2021,Kuhlenkamp24}. In the following, we refer to this additional flavor simply as spin. At the critical point of the topological transition, the charge gap collapses, while the spin sector remains gapped throughout both the FCI phase and the semion crystal reached by strengthening repulsive interactions. This quantum criticality thereby enforces a hierarchy in which the lowest-energy charge carriers are spin singlets. Near criticality, therefore, anyons with {\em even} numerator charge cost less energy, guaranteeing that the cheapest anyons fuse to a spinless boson (eg. a Cooper pair) rather than to a spinful electron. Doping either side of the critical point over a broad parameter regime should thus yield charge-2e superconductivity, possibly coexisting with translation-symmetry breaking inherited from the crystal (see Fig.~\ref{fig:overview}).

{\em Concrete realization and main results:}  We consider realizing spin-singlet Halperin FQH states in flat Chern bands.  
The Halperin states are labeled by the integers $(mmn)$~\cite{Halperin:1983zz} where for electrons $m$ is an odd integer. Further, we focus on spin-singlet states that arise in the special case of $m=n\pm1$, and $n\in2\mathbb{Z}$~\cite{FradkinSinglet1991,Jain2007,DongSinglets}. We discuss these states at general fillings of the Chern band $\nu = \frac2{2n\pm 1} $ with a parton-vortex approach, within which we can describe the FCI, the semion crystal, the superconductor, and the transitions between them.

Focusing on $\nu=2/3$, we demonstrate a robust FCI using density-matrix renormalization‐group simulations of a purely repulsive microscopic model. We further demonstrate that increasing Hubbard repulsion drives the FCI into the semion crystal---a charge-density wave of electrons supporting an emergent chiral spin liquid. The fact that the tuning of a simple repulsive interaction creates this exotic topological crystal shows that it can form easily despite its complexity and that it is likely to be a relevant competing phase to the FCI. Moreover, the Cooper pair-field correlations {\em peak} in the vicinity of the transition, signaling incipient superconductivity.  

We further discuss how the mechanism generalizes to fillings $\nu = 2/(2n\pm 1)$. Although $\nu=2/3$ enjoys special simplifications, the same logic of the energetic hierarchy applies: spin singlet, bosonic charge-2e anyons emerge as the preferred low-energy carriers near the corresponding topological transitions. The transition between the FCI and SX can be re-expressed as a plateau transition of Cooper pairs from a hierarchy state $\nu_{\rm Cooper}= 1/(\mp 2- n^{-1})$ to $\nu_{\rm Cooper}=\mp1/2$, providing an intuitive bosonic description of all three phases—FCI, semion crystal, and superconductor—and clarifying their interrelations.  Interestingly, the relative sign of the charge Hall conductance in the FCI and the sense of propagation of the chiral edge mode in the semion crystal alternates with the $\pm 1$ choice in the fractions above. Doping both the FCI and the semion crystal near the transition favors the formation of superconductors, which in turn inherit the alternating chirality of the semion crystal. At these more general fillings, we find that a generic pathway to charge-2e superconductivity without residual topological order exists---provided the superconductor spontaneously breaks translation symmetry. Translationally invariant charge-2e superconductors can also appear upon doping FCIs, yet within our analysis these phases retain an additional topological sector and should be identified as SC$^*$ states, with $\nu = 2/3$ being the exception. 

Our analysis, therefore, unifies several recent proposals for anyon superconductivity from studies of lattice models to theories for superconductivity proximate to FQAH states in twisted MoTe$_2$. Previous studies of the Hubbard–Hofstadter platform focused on the integer filling $\nu = 2$ (corresponding to $m = 1, n = 0$), where  topological criticality, chiral spin liquids and superconductivity were identified ~\cite{Kuhlenkamp24,KuhlenkampThesis, Divic24,divic_anyonsuperconductivity_2024}. Here we venture into genuinely fractional fillings, where fractional quantum Hall states emerge, and investigate how the topological criticality mechanism manifests in this regime. 

Our work is structured as follows: We provide a field-theoretic analysis of the transition between singlet FQH states and a topological CDW, based on a parton description, in Sec.~\ref{sec:Transition}, and elaborate on the mechanism for anyon superconductivity in the vicinity of the critical point in Sec.~\ref{sec:Mechanism}. Using the analytical insights, we then develop a microscopic Hubbard-Hofstadter model and numerically study its phase diagram at $\nu=2/3$ using tensor network techniques in Sec.~\ref{sec:Numerics}. Our results are consistent with a continuous topological transition between a bilayer FQH phase and the semion crystal. Crucially, we also numerically confirm that across the transition the spin gap remains open while the charge gap closes. We discuss experimental implications, including potential consequences for recent experimental observations in twisted MoTe$_2$, in Sec.~\ref{sec:ExperimentalImplications} and provide an outlook and discuss the general insights into anyon superconductivity obtained from our formalism in Sec.~\ref{sec:Outlook}.

\section{Phases and Phase transitions at fixed filling: Fractional Chern Insulator and semion crystal}
\label{sec:Transition}

\begin{figure}
\begin{center}
\includegraphics[width=0.99\linewidth]{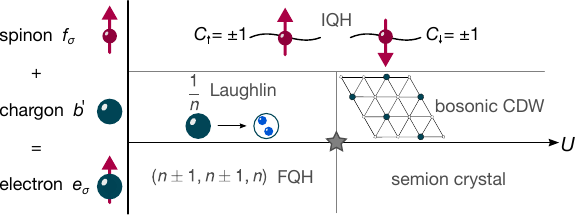}
\caption{\textbf{Parton construction.}  Singlet FQH states can be described using fermionic spinons in IQH states at level $C_{f_\sigma}=\pm1$ and bosonic chargons in a $1/n$ Laughlin state. To describe a continuous transition to a semion crystal, the chargons undergo a crystallization transition, forming a bosonic CDW. The spinons remain in the same state across the transition, implying that the spin gap stays finite.}
\label{fig:partonconstruction}
\end{center}
\end{figure}

In this section, we show that favorable conditions for anyon superconductivity emerge in isolated Chern bands that host a (pseudo-)spin flavor, such as real spin, a layer, or valley index. A microscopic mechanism for forming anyon superconductivity is then provided when strong repulsion drives these fractionally filled states near competing topological CDW states. Based on parton constructions, we develop a field-theoretic description of such a transition from a Fractional Chern Insulator to the semion crystal (describing a CDW state atop of which a chiral spin liquid forms). This generalizes previous results on the topological transition between an integer quantum Hall state and a chiral spin liquid~\cite{Kuhlenkamp24,KuhlenkampThesis,Divic24,divic_anyonsuperconductivity_2024} to fractional fillings. {While we focus on Halperin states in the following, our arguments can be generalized to the Jain sequence (see Appendix~\ref{app:Jain_sequence}).}

\subsection{Fractional Chern insulators}
\label{sec:parton_theory_fci}
At a total filling of $\nu = \nu_\uparrow+\nu_\downarrow = 2/(m+n)$, interactions can stabilize Halperin FQH states, labeled by the integers $(mmn)$~\cite{Halperin:1983zz}, where $m$ is odd for fermionic states. The low-energy properties of these states are described by topological quantum field theories with the following effective Lagrangian
\begin{equation}
\begin{aligned}
\mathcal{L}_\mathrm{FCI} = &-\frac{1}{4\pi}\sum_{I,J}K_{IJ} \alpha_I \wedge \; \mathrm{d}\alpha_J + \frac{1}{2\pi}\sum_I q_I^c\alpha_I \wedge \mathrm{d}A \\
&+ \frac{1}{2\pi}\sum_Iq^s_I  \alpha_I \wedge \mathrm{d}A_s   ,
\end{aligned} \label{eq:AbelianFQHLagrangian}
\end{equation}
where $\alpha_I \wedge \mathrm{d} \alpha_J =  \epsilon_{\mu \nu \lambda} \alpha_{I,\mu} \partial_\nu \alpha_{J,\lambda}$ and the $K$-matrix, charge vector $\mathbf{q}^c$ and spin vector $\mathbf{q}^s$ are defined as
\begin{equation}
         K = \begin{pmatrix}
        m & n \\
        n  & m
\end{pmatrix}, \quad \mathbf{q}_c = (1,1)^T, \quad \mathbf{q}_s = (1, -1)^T.
\label{eq:mmn-Kmatrix}
\end{equation}
Furthermore, $A$, $A_s$ represent the background electromagnetic fields that keep track of charge and spin, respectively. $A_s$ is a background gauge field that couples to spin $S_z$, which we introduce to keep track of the spin degree of freedom.\footnote{For pseudo spins which are encoded as a layer-degree of freedom, $A_s$ may well be physical and would, for instance, dictate the counterflow response.} For fermions, $m$ must be an odd integer. Among the fermionic Halperin states $(mmn)$, the sequence with $m=n\pm1$, at filling $\nu=\frac{2}{2n\pm1}$ where $n$ is an even integer, are of particular interest here. These states enjoy full spin rotation invariance, and furthermore, the $(112)$ state with filling $\nu=2/3$ has the same topological order as the $\nu=2/3$ Jain state, which is commonly observed in experiments of spin-polarized systems~\cite{Tsui1982,cai_SignaturesFractional_2023, zeng_ThermodynamicEvidence_2023, park_ObservationFractionally_2023, xu_ObservationInteger_2023}, while the $\nu=2/5$ Halperin state has the same topological order as the commonly observed $\nu=2/5$ Jain state. Although we construct our theory for Halperin states for which our arguments become particularly transparent,  and the presence of spin symmetry clarifies some of our arguments, the concepts we illustrate are applicable much more broadly, as we also discuss in Sec.~\ref{sec:ExperimentalImplications}.

Quasiparticles of states in the sequence $m=n\pm1$ are labeled by an integer vector $\mathbf{l} = (l_\uparrow, l_\downarrow)^T$, which denotes the couplings of the quasi-particle current to the emergent gauge fields. The charges and self-statistics of an anyon $\mathbf{l}$ are given by:
    \begin{equation}
    \begin{aligned}
    Q^\mathbf{l}&= -e \mathbf{q}_c^T K^{-1}\mathbf{l} = - e \frac{l_\uparrow+l_\downarrow}{2n\pm 1}\\
     S^\mathbf{l}_z & =  \mathbf{q}_s^T K^{-1}\mathbf{l} =  \pm (l_\uparrow-l_\downarrow) \\
     \theta^\mathbf{l} & = \pi \mathbf{l}K^{-1}\mathbf{l} = \frac{\pi}{2n\pm 1}[\mathbf{l}^2(1\pm n) \mp2n\, l_\uparrow l_\downarrow ].
     \end{aligned}
    \end{equation}
We find that the $\mathbf{l}_\uparrow = (1,0)^T$ and $\mathbf{l}_\downarrow = (0,1)^T$ anyons have electric charge $-e/(2n\pm1)$. These anyons with the smallest electric charge carry the same spin $S_z=\pm 1/2$ quantum numbers as the original electrons, while $\mathbf{l}_2 = (1,1)^T$ anyons carry vanishing spin $S_z = 0$ and have twice the charge $Q=-2e/(2n\pm1)$. Thus, the electron is decomposed of anyons which do not carry a fractionalized spin and the original action of an electronic $\mathrm{SU(2)}$ spin symmetry is preserved on the level of the anyons. This is a special property of the $(n\pm1,n\pm1,n)$ states consistent with their spin-singlet nature~\cite{FradkinSinglet1991, mcdonaldTopologicalPhaseTransition1996}.

%, which is a special feature of the $(n\pm1,n\pm1,n)$ states%, as it makes their topological order compatible with an $\mathrm{SU(2)}$ spin rotation symmetry~\cite{FradkinSinglet1991}. On the other hand, $\mathbf{l}_2 = (1,1)^T$ anyons carry vanishing spin $S_z = 0$ and have twice the charge $Q=-2e/(2n\pm1)$. 

We now develop a parton theory for these singlet FQH states, which is naturally suited for studying topological phase transitions to competing phases; see Fig.~\ref{fig:partonconstruction}.

Since the spin remains unfractionalized in the Halperin states, they can be conveniently described using a parton description, where we decompose the electron operator $e_\sigma$ into a fermionic {\em spinon} $f_\sigma$, which carries its spin, and a {\em chargon} $b'(x)$
\begin{equation}
     e_\sigma(x) =   f_\sigma(x) b'(x).
\label{eq:partons_fb}
\end{equation}
Written in this way $b'(x)$ is a hard-core boson, which is appropriate for our setting, as we are working at fractional filling of the electrons. The decomposition of the electron into spinon and chargon leads to a gauge constraint
\begin{equation}
n_e(x) \equiv \sum_\sigma e^\dagger_\sigma e_\sigma(x) = \sum_\sigma f_\sigma^\dagger f_\sigma(x) = b'^\dagger b'(x),   \label{eq:gauge_constraint}
\end{equation}
which is enforced by $f_\sigma(x)$ and $b'(x)$ carrying opposite gauge charge, see Eq.~\eqref{eq:partons_fb}.

To illustrate how the partons encode the fractional quantum Hall states, it is convenient to consider a lattice system in a background magnetic field giving rise to $\Phi = \Phi_0/2$ external flux per unit cell of an underlying lattice, where $\Phi_0$ is the flux quantum. To find $(n\pm 1,n\pm1,n)$-type quantum Hall states, the total filling of the Chern bands must be $\nu=2/(2n\pm1)$, which implies that the electron density is $n_e =\nu \Phi/\Phi_0 =1/(2n\pm1)$ per unit-cell.\footnote{While it is convenient to consider finite background flux; it is straightforward to adjust the construction to the pure FCI setting without background magnetic field $B_z=0$.} Due to the constraint, spinons are at density $n_{f_\uparrow}=n_{f_\downarrow}=n_e/2$ and we assume that both spinons fill Chern bands with $C_{f_\sigma}=\pm 1$, such that they are in an integer quantum Hall (IQH) state with total Chern number $C_f=\pm 2$. The parton construction Eq.~\eqref{eq:partons_fb} can be understood as a form of flux attachment, such that the flux seen by each parton partially cancels the external flux. From the Streda formula for the integer quantum Hall state of the spinons, we require that the internal gauge flux seen by them is $\Phi_f = \Phi_0(n_{f_\uparrow}+n_{f_\downarrow})/C_f$. Consequently, the remaining flux $\Phi_{b'}$ seen by the chargons is
\begin{equation}
    \Phi_{b'} = \Phi - \Phi_f =  \frac{n}{2n\pm1} \Phi_0,
    \label{eq:fluxChargon}
\end{equation}
Now, since the chargons are at a density:
\begin{equation}
    n_{b'} = n_e = \frac{1}{2n\pm 1},
\end{equation}
they can naturally form a bosonic Laughlin state at filling
\begin{equation}
        \nu_{b'} = n_{b'}/\Phi_{b'}= 1/n.
\end{equation}
To obtain a commensurate flux for all fillings, the spinon Chern number $C_f=\pm2$ had to alternate in sign.
We remark that in the trivial case, $n=0$ at a density $n_e=1$, all the flux is absorbed by the spinons, see Eq.~\eqref{eq:fluxChargon}, and the chargons see zero net flux. This case was analyzed previously in~\cite{Kuhlenkamp24,divic_anyonsuperconductivity_2024} and we focus here on the non-vanishing $n$.

We can now show that this correctly describes the $(mmn)=(n\pm1,n\pm1,n)$ Halperin state. The total Lagrangian of the electronic system is given by
\begin{equation}
\mathcal{L}_\mathrm{FCI}= \mathcal{L}_f[a] + \mathcal{L}_{b'}[A-a].
\end{equation}
The spinons contribute the following term
\begin{equation}
    \mathcal{L}_f[a] = \mp \frac{1}{4\pi} ( \alpha_\uparrow \wedge \mathrm{d}  \alpha_\uparrow + \alpha_\downarrow \wedge \mathrm{d}  \alpha_\downarrow) + \frac{1}{2\pi} a \wedge \mathrm{d}(\alpha_\uparrow + \alpha_\downarrow)
\end{equation}
while the chargons in the $1/n$ bosonic Laughlin state are described by
\begin{equation}
    \mathcal{L}_{b'}[A-a] = -\frac{n}{4\pi} \beta \wedge \mathrm{d}  \beta + \frac{1}{2\pi} \beta \wedge \mathrm{d}(A - a). \label{eq:bosonicLaughlin}
\end{equation}
The spinon currents $j^\mu_{(\sigma)} = 1/2\pi \varepsilon^{\mu \nu \lambda} \partial^\nu \alpha_{(\sigma)}^\lambda$, are expressed in terms of the gauge fields $\alpha_\sigma$, with $\sigma \in \{\uparrow, \downarrow\}$. The additional gauge field $\beta$ describes the bosonic Laughlin state. Strictly speaking, in order to obtain a K-matrix that satisfies the spin-charge relation, which stems from the fact that in electronic condensed matter systems the only unfractionalized charge $e$ excitation is the electron, a fermion, we need to couple the external gauge field with odd (even) charge to fermionic (bosonic) partons. This is achieved by shifting $\tilde{a}\rightarrow \left (a-A \right)$ and simplifying notation with $\tilde{a}\rightarrow a$. Then the final Lagrangian is:
\begin{eqnarray}
\mathcal{L}_\mathrm{FCI}&=& \mp \frac{1}{4\pi} ( \alpha_\uparrow \wedge \mathrm{d}  \alpha_\uparrow + \alpha_\downarrow \wedge \mathrm{d}  \alpha_\downarrow) 
    \nonumber\\ &&+\frac{1}{2\pi} \left (a+A\right ) \wedge \mathrm{d}(\alpha_\uparrow + \alpha_\downarrow)
   \nonumber\\ 
   && -\frac{n}{4\pi} \beta \wedge \mathrm{d}  \beta - \frac{1}{2\pi} \beta \wedge \mathrm{d}a 
    . \label{eq:FullParton}
\end{eqnarray}

Now, the $a$ are conventional gauge fields, and recall that $n$ is even. Integrating out $a$ enforces the constraint $\beta = \alpha_\uparrow + \alpha_\downarrow$ and we obtain Eq.~\eqref{eq:AbelianFQHLagrangian}, with K-matrix Eq.~\eqref{eq:mmn-Kmatrix} for $m=n\pm1$. The two different signs are understood to arise from spinons forming Chern bands with $C=\pm 1$.
For states of the form $(n-1, n-1,n)$, the chirality of the spinons is opposite to that induced by the background flux, resulting in counter-propagating spin and charge edge modes, which is why they are commonly referred to as $(\overline{n-1},\overline{n-1},n)$ states.

\subsection{Semion crystal}
A competing class of gapped ground states at fractional fillings are charge density waves (CDWs), stabilized by minimizing the long-range Coulomb repulsion. One option to undergo a crystallization transition is to simultaneously close the charge and spin gap of the $(n\pm1,n\pm1,n)$ FQH state, resulting in a trivial crystal, most likely via a first-order transition. 

More interestingly, a crystal can also form by only closing the charge gap, activating only the bosonic chargons. Such crystallization transitions of bosonic degrees of freedom can be continuous~\cite{Balents_puttingorders_2005,Balents2005, Song2024}. This competing crystalline order, which we refer to as semion crystal in the following, the transition into it, and the proximate phases on doping is the focus of our work.
The properties of this state are particularly transparent in our parton construction. There, we decompose the electron into the spinon and hardcore chargon $e_\sigma(x) = f_\sigma(x) b'(x)$,
where the bosonic parton $b'(x)$ is at {\em fractional} filling of the underlying lattice with density $n_b<1$. Long-range interactions can then lead to a crystallization transition at which the boson gap closes, while the spin gap, and therefore the gap of the fermionic partons $f_\sigma$, remains open. Once the $b'(x)$ form a crystal, these degrees of freedom are trivially gapped and can be integrated out. To leading order, they only leave behind a fluctuating gauge field $a$, and, after coupling to the external spin gauge field we have:
\begin{equation}
\begin{aligned}
    \mathcal{L}_{\text{SX}} [a,A_s] =& \mp \frac{1}{4\pi} ( \alpha_\uparrow \wedge \mathrm{d}  \alpha_\uparrow + \alpha_\downarrow \wedge \mathrm{d}  \alpha_\downarrow) \\
    +\frac{1}{2\pi} a \wedge& \mathrm{d}(\alpha_\uparrow + \alpha_\downarrow) + \frac{1}{2\pi} A_s \wedge \mathrm{d}(\alpha_\uparrow - \alpha_\downarrow) 
\end{aligned}
\end{equation}
which, after integrating out $a$, further simplifies the field theory of the crystal to 
\begin{equation}
\begin{aligned}
    &\mathcal{L}_{\text{SX}} 
        =\mp \frac{2}{4\pi}\alpha\wedge \mathrm{d}\alpha + \frac{2}{2\pi} A_s \wedge \mathrm{d}\alpha,
        \label{eq:semion_crystal}
\end{aligned}
\end{equation}
where $\alpha_\uparrow =-\alpha_\downarrow =: \alpha$. We see that the crystal is an electrical insulator with vanishing Chern number (no dependence of $A$), and describes a chiral spin liquid (CSL) with semion topological order, which we refer to as semion crystal. 
The semion crystal can be described from two complementary perspectives. At long wavelengths, the pictures coincide, but their relative utility depends on how deep one is in the ordered phase. Focusing on the bosonic operators $b'(x)$ that crystallize, we note that they carry unit electric charge and acquire semionic statistics through their coupling to the emergent gauge field. The phase can therefore be viewed as a CDW of spinless semions. Moving further into the crystalline phase, the charge gap grows while the super-exchange processes that generate spin dynamics are progressively suppressed. In this limit, the most transparent description is of a chiral spin liquid formed from the spins of the frozen electronic CDW.

The alternating quantum spin Hall response seen in Eq.~\eqref{eq:semion_crystal} is inherited from the spin modes of the Halperin states. This is discussed in more detail in Sec.~\ref{sec:Numerics}, where we study a microscopic model.

\subsection{Interpretation of the FCI to semion crystal transition as bosonic Cooper pair transition}
\label{sec:EffecTheoryCooperPairs}

Here we point out that transitions like $(\bar{1}\bar{1}2) \leftrightarrow \text{semion crystal}$ can be interpreted as {\em purely bosonic} plateau transitions of Cooper pairs. This immediately suggests that there are Cooper pairs near the transition. A similar argument was made for the integer quantum Hall (with $\sigma_{xy}=2 e^2/h$) to CSL transition in Ref.~\cite{divic_anyonsuperconductivity_2024}. Here, we show that a related picture emerges for all the fractional electronic states we consider in this paper as well.

To illustrate this, consider the transition from the $(\bar{1}\bar{1}2)$ FQH state with $\lbrace \sigma_{xy},\sigma_{xy}^s\rbrace =\lbrace 2/3,-2\rbrace $ to the semion crystal with $\lbrace \sigma_{xy},\sigma_{xy}^s\rbrace =\lbrace 0,-2\rbrace $. We can trivialize the spin response by stacking a $\nu_{\text{tot}}=2$ IQH state on top of it, composed of spin up and spin down electrons, each in a Chern band with unit Chern number. Since these are integer quantum Hall layers with no bulk anyons, their only effect is to change the edge physics, and hence bulk properties of the original system are unchanged upon stacking. After incorporating the IQH state, this gives: $\lbrace \sigma_{xy},\sigma_{xy}^s\rbrace =\lbrace 8/3,0\rbrace $ for the FQH and $\lbrace \sigma_{xy},\sigma_{xy}^s\rbrace =\lbrace 2,0\rbrace $ for the semion crystal. Note, the spin response has been trivialized, which allows for a description of the transition purely in terms of spinless degrees of freedom (see Appendix~\ref{app:QCP_CooperPairs} for more details). It is nothing but a plateau transition between a Laughlin state of spin singlet Cooper pairs with $\sigma_{xy}^\text{Cooper}=\frac{1}{2}\frac{(2e)^2}{h}=2\frac{e^2}{h}$ and its hierarchy state $\sigma_{xy}^\text{Cooper}=(\frac{1}{2}+\frac{1}{6})\frac{(2e)^2}{h}=\frac{2}{3}\frac{(2e)^2}{h}=\frac83\frac{e^2}{h}$. 

The nature of this plateau transition can be understood as follows. Consider a parent state built of the Laughlin state of Cooper pairs with an excess filling of $\nu^{\text{Cooper}}_{\text{excess}}=1/6$ of semions. Two options then become available. First, the excess density of semions could crystallize, breaking translation symmetry and leaving behind semion topological order. Or, they could form their own FQH state, which is the basis of the hierarchy construction.

This argument can be generalized to all other abelian spin singlet states\footnote{The relevant hierarchy states of Cooper pairs appear at filling $\nu_{\text{Cooper}}=\mp\frac{1}{2}+ \frac{1}{4n\pm 2 }$,} at filling $\nu = 2/(2n \pm 1)$ which have $\lbrace \sigma_{xy},\sigma_{xy}^s\rbrace =\lbrace 2/(2n\pm1), \pm 2 \rbrace$. 
Again, we must incorporate the spin singlet IQH state, with alternating sign of the Hall conductance, to cancel the spin response. Doing this gives: $\lbrace \sigma_{xy},\sigma_{xy}^s\rbrace =\lbrace \mp2,0\rbrace $ for the semion crystal and $\lbrace \sigma_{xy},\sigma_{xy}^s\rbrace =\lbrace \frac2{2n\pm1} \mp 2,0\rbrace $ for the FQH state. Again, this is nothing but a plateau transition between a Laughlin state of spin singlet Cooper pairs (Hall conductance $\sigma_{xy}^\text{Cooper}=\mp\frac{1}{2}\frac{(2e)^2}{h} $), and a hierarchy state with Hall conductance: $\sigma_{xy}^\text{Cooper}=(\frac{1}{\mp 2-1/n})\frac{(2e)^2}{h}$, where we recall that $n$ is even.

\subsection{Effective theory of the critical point}
\label{sec:EffectiveTheoCritPoint}
\begin{table}[]
    \centering
        \begin{tabular}{l|cccc|c}
        \toprule
         & $a_1$ &  $a_2$ & $a$ & $A$ & $C$\\
        \hline
        $f_\sigma$ & 1 &  0 &0 & 0& $\pm2$ \\
        $g_1$ & -1 & 1 & 0&0 & 1 \\
        $g_2$ & 0 &  -1  & 1&0& 1 \\
        $b$ & 0 & 0 &-1 & 1 &\\
        \bottomrule
        \end{tabular}        
        \caption{List of partons, their charge under the emergent gauge fields $a_{1,2}$, the background field $A$, and their total Chern number $C$ for $\nu=2/(4\pm1)$.} \label{tab:redFermionicPartons}
\end{table}

In the previous sections, we gave a heuristic description of the transition between the FQH and semion crystal phases, first in terms of the bosons $b'$ transitioning between a bosonic Laughlin and a crystalline insulating phase, and later as a plateau transition between bosonic hierarchy states of Cooper pairs. We now derive the effective theory of the critical point in the former picture using the parton construction. There the bosons, as described in Eq.~\eqref{eq:partons_fb}, are both at a finite density and see a background flux. To obtain a convenient effective theory, we would like instead to work with bosonic variables that effectively are at zero density, represented by a rotor field. Note, simply going over to the vortex description of $b'$ is not sufficient in this case since it also experiences a flux, and hence the vortices are at finite density. Instead, we will introduce an extended parton construction, which will help us define this new bosonic field. We illustrate this for the simple case of $n=2$ corresponding to $\nu=2/3$ and $2/5$, where the extended construction is:
 \begin{equation}
 e_\sigma(x) = f_\sigma g_1g_2b(x).
 \label{eq:partons_fg1g2b}
 \end{equation}

The fermions $g_{1,2}$ are at density $n_b=n_{g_1} = n_{g_2}$ and couple to two additional emergent gauge fields $a_{1,2}$ as outlined in Table~\ref{tab:redFermionicPartons}. Our system is then described by the following Lagrangian
\begin{equation}
    \mathcal{L}= \mathcal{L}_{f_\sigma}[a_1]+\mathcal{L}_{g_1}[a_2-a_1] + \mathcal{L}_{g_2}[a-a_2]+ \mathcal{L}_b[A-a].
\label{eq:extended_parton_lagrangian}
\end{equation}
The background fluxes $\nabla\times a_{1,2}$ and $\nabla\times a$ can now be adjusted such that the fermions $\lbrace f_\sigma,g_1,g_2\rbrace$ form IQH states at fillings $\lbrace -2,1,1 \rbrace$ for $\nu=2/3$ and $\lbrace 2,1,1 \rbrace$ for $\nu=2/5$. At the same time, this choice ensures a zero-flux condition for the bosons $\nabla\times (A-a) =0$. Once the fermions are in gapped IQH states, they contribute terms of the form
\begin{equation}
\begin{aligned}
\mathcal{L}_{f_\sigma}[a_1] &=\pm  \frac{1}{4\pi} ( \alpha_{\sigma} \wedge \mathrm{d}  \alpha_{\sigma})+ \frac{1}{2\pi} a_1 \wedge \mathrm{d}\alpha_\sigma\\
    \mathcal{L}_{g_i}[a'] &=- \frac{1}{4\pi} ( \beta_{i} \wedge \mathrm{d}  \beta_{i})+ \frac{1}{2\pi} a' \wedge \mathrm{d}\beta_i;
    \end{aligned}
\label{eq:extended_parton_lagrangian_contd}
\end{equation}
hence, we find that Eq.~\eqref{eq:extended_parton_lagrangian} takes a simple form. We only quote the result for general $n$\footnote{for general fillings $\nu=\frac{2}{2n\pm1}$ we introduce $n$ additional fermions (recall n is even), see Appendix~\ref{app:generalCharge2SC}.}:
\begin{equation}
\begin{aligned}
     \mathcal{L}  = &\mathcal{L}_b[a-A] - \frac{n}{4\pi} \beta \wedge \mathrm{d}\beta + \frac{1}{2\pi} \mathrm{d}\beta \wedge (a-a_1 ) \\
     \mp&
       \frac{1}{4\pi}(\alpha_\uparrow \wedge \mathrm{d}  \alpha_\uparrow + \alpha_\downarrow \wedge \mathrm{d}  \alpha_\downarrow) \\
     +&  \frac{1}{2\pi}a_1\wedge \mathrm{d}(\alpha_\uparrow +\alpha_\downarrow)+\frac{1}{2\pi}A_s \wedge \mathrm{d}(\alpha_\uparrow -\alpha_\downarrow).
\end{aligned}
\end{equation}
Thus, in this extended picture we can now analyze the transition in terms of a standard superfluid to a CDW transition of the bosons in the absence of background flux. {As before, the FQH state forms when $b$ condenses. Close to the transition, vortices of the $b$ condensate are natural excitations, which become soft as the transition is approached. Thus, to describe excitations in the FQH state it is convenient to work directly with vortex operators. This is achieved by going} to a dual vortex formulation of $b(x)$~\cite{lannert2001, Balents2005, Balents_puttingorders_2005}. Writing the boson current as $j_b^\mu = 1/2\pi\epsilon^{\mu\nu\sigma}\partial_\nu \beta_\sigma$
, boson-vortex duality tells us that the vortex field operator $\tilde{\varphi}(x)$ couples minimally to $\beta$. When bosons are at a nonzero background density, the equations of motion tell us that there is a background flux $\mathrm{d}\beta/2\pi$ equal to the background electron density. This, in turn, leads to $p=2n\pm1$ valleys {in the disperison of} the bosonic vortex field.
The Lagrangian describing the phase transition reads:
\begin{equation}
\begin{aligned}
     \mathcal{L}_{\mathrm{QCP}}  &= |D_{\beta}\tilde{\varphi}|^2 - V_\text{int}(\tilde{\varphi}) - \frac{n}{4\pi} \beta \wedge \mathrm{d}\beta \\
     -&\frac{1}{2\pi} \mathrm{d}\beta \wedge a 
     \mp  \frac{1}{4\pi}(\alpha_\uparrow \wedge \mathrm{d}  \alpha_\uparrow + \alpha_\downarrow \wedge \mathrm{d}  \alpha_\downarrow) \\
     +& \frac{1}{2\pi} (a+A) \wedge  \mathrm{d}(\alpha_\uparrow + \alpha_\downarrow) + \frac{1}{2\pi}A_s \wedge \mathrm{d}(\alpha_\uparrow -\alpha_\downarrow).
     \label{eq:L_QCP}
\end{aligned}
\end{equation}
The vortex theory can have an emergent Lorentz symmetry, as the condensation transition is not density-changing. As we have seen for a fractional boson density $n_b =n_e= q/p$, the vortices see $\Phi_v =2\pi q/p$ flux per unit cell\footnote{we have $q=1$ for $\pi$ flux per unit cell and $q=2$ for FCIs without background flux.}, where $p=(2n\pm 1)$ is equal to the number of valleys. Consequently, in the low energy description, there are $p$ vortex fields $\tilde{\varphi} = (\tilde{\varphi}_1, \dots,\tilde{\varphi}_p)$, which are related under the magnetic translation symmetries~\cite{Balents2005} by the transformation $T_1:\tilde{\varphi}_j\rightarrow \tilde{\varphi}_{j-1}$ and $T_2:\tilde{\varphi}_j\rightarrow \omega_v^j\tilde{\varphi}_{j}$, where the phase factor is $\omega_v = e^{i2\pi \Phi_v}$. Expressing Eq.~\eqref{eq:L_QCP} in terms of the low-energy vortex fields and simplifying the Chern-Simons part by integrating out $a$, leads to the effective theory describing the FCI and SX phases, and their transition:
\begin{equation}
\begin{aligned}
     \mathcal{L}_{\mathrm{QCP}}  = \sum_{j=1}^p &\left \{|D_{\alpha^{(1)}_\uparrow + \alpha^{(1)}_\downarrow}\tilde{\varphi}_j|^2 - s|\tilde{\varphi_j}|^2\right \} - V_{\text{int}}(\tilde{\varphi}_j)\\ - \frac{n}{4\pi} (\alpha_\uparrow + \alpha_\downarrow) \wedge &\mathrm{d}(\alpha_\uparrow + \alpha_\downarrow) 
     \mp  \frac{1}{4\pi}(\alpha_\uparrow \wedge \mathrm{d}  \alpha_\uparrow + \alpha_\downarrow \wedge \mathrm{d}  \alpha_\downarrow) \\
     + \frac{1}{2\pi} A \wedge  \mathrm{d}(\alpha_\uparrow &+ \alpha_\downarrow) + \frac{1}{2\pi}A_s \wedge \mathrm{d}(\alpha_\uparrow -\alpha_\downarrow),
     \label{eq:L_QCP_effective}
\end{aligned}
\end{equation}
 We note that the effective low-energy theory is just bosonic QED$_3$+Chern-Simons, with $p$ flavors of bosons, coupled to a gauge field with a Chern-Simons action. Since the background flux of $\alpha_{\uparrow,\downarrow}$ is already taken into account by the flavor index, vortices couple only to fluctuations of the gauge fields $\alpha^{(1)}_{\uparrow,\downarrow} = \alpha_{\uparrow ,\downarrow} - \langle \alpha_{\uparrow ,\downarrow} \rangle$. We can access the two phases by varying the boson mass $s$, while $V_{\text{int}}$ represents interactions between vortex fields that are constrained by the magnetic symmetry. 

Let us now confirm that the vortex theory correctly reproduces the two insulating phases. 
When $s>0$ and all the vortex fields $\tilde{\varphi}$ are gapped, we obtain the effective action: 
\begin{equation}
\begin{aligned}
     \mathcal{L}_\mathrm{FCI} = &- \frac{n}{4\pi} (\alpha_\uparrow + \alpha_\downarrow) \wedge \mathrm{d}(\alpha_\uparrow + \alpha_\downarrow) 
     \\&\mp  \frac{1}{4\pi}(\alpha_\uparrow \wedge \mathrm{d}  \alpha_\uparrow + \alpha_\downarrow \wedge \mathrm{d}  \alpha_\downarrow) \\
     &+ \frac{1}{2\pi} A \wedge  \mathrm{d}(\alpha_\uparrow + \alpha_\downarrow) + \frac{1}{2\pi}A_s \wedge \mathrm{d}(\alpha_\uparrow -\alpha_\downarrow),
\end{aligned}
\end{equation}
which is nothing but the action for the FQH states, i.e.  Eq.~\eqref{eq:AbelianFQHLagrangian}, with 
\begin{equation}
    K = \begin{pmatrix}
        n\pm1 & n \\
        n & n\pm1
    \end{pmatrix}.
\end{equation}
The transition to the crystal is described by the {\em condensation} of the vortex field $\tilde{\varphi}$, when $s<0$, which leads to a Meissner effect for the internal gauge field, setting $\alpha_\uparrow =- \alpha_\downarrow =\alpha$ and reducing to Eq.~\eqref{eq:semion_crystal}. In this way, the coupling to the external gauge field $A$ is lost, signaling the formation of an insulator 
 while retaining semion topological order and quantized spin transport. In the process of vortex condensation, we need to pick out one combination of vortex fields to condense, which breaks translation symmetry. Thus, in this limit, our theory describes a semion crystal. Note, we obtain essentially the same topological order for all values of $p$ in the SX phase, except for the $\pm$ dependence in Eq.~\eqref{eq:semion_crystal} on whether $p=2n\pm 1$, in contrast to the topological order of the FQH state that depends strongly on $p$. 

The critical point is described by the theory obtained on setting $s=0$, which at this level of approximation describes a continuous transition. Both the change of topological order and translation symmetry breaking occur simultaneously on crossing the critical point. This occurs because the $p=(2n\pm 1)$ flavors of vortices that undergo the condensation transition carry nontrivial statistics ($\theta_v = \pi \nu$)
  that are related to the  filling $\nu=\frac{2}{2n\pm1}$ of the FQH phase. Note, in the limit $n\rightarrow \infty$, the statistics is trivialized, and one approaches the usual complex scalar field transition with $p$ components.

We emphasize that the transition from spin-singlet Halperin states to the semion crystal phase is driven by the crystallization of the chargons (dual to the vortex fields described above), while the spin sector remains unchanged. This implies that the spin gap remains open across the transition, energetically favoring doping of spin-singlet excitations. Close to the critical point, the vortex field is soft and its current couples with vector $\mathbf{l}=(1,1)^T$ to $\alpha_\uparrow,\alpha_\downarrow$, resulting in vortices carrying charge
\begin{equation}
    Q_{\tilde{\varphi}} = -\frac{2e}{2n\pm1}.
\end{equation}
Charge fluctuations at the critical point are then governed by vortices, or equivalently by spinless anyons. Microscopic details encoded in the vortex interactions $V_\text{int}(\tilde{\varphi})$ determine the pattern of vortex field condensation $\tilde{\varphi}_i$, allowing for the formation of different types of charge density waves~\cite{Balents2005}. 
The special case of the bilayer {\em integer} quantum Hall $(110)$ to CSL transition, as discussed in Refs.~\cite{Kuhlenkamp24, KuhlenkampThesis, Divic24, divic_anyonsuperconductivity_2024}, can be understood from setting $n=0$ within the broader framework developed here. There too superconductivity emerges on doping the topological critical point~\cite{divic_anyonsuperconductivity_2024}, although all Hall conductances are integral. Here we will see that, surprisingly, an analogous outcome is realized even when the insulators exhibit {\em fractional} Hall conductance.

\section{Finite Doping: Superconductivity} \label{sec:Mechanism}

The existence of a spin gap at the quantum critical point, while the charge gap softens, provides the foundation for anyon superconductivity and rationalizes the equivalent plateau transition of Cooper pairs. On the semion crystal side of the transition, the energetic separation of spinful and spinless excitations provides a natural mechanism for superconductivity upon doping the Kalmeyer-Laughlin state~\cite{kalmeyer87}, which dates back to seminal work by Laughlin~\cite{laughlinRelationshipHighTemperatureSuperconductivity1988, laughlinSuperconductingGroundState1988, fetterRandomphaseApproximationFractionalstatistics1989, leeAnyonSuperconductivityFractional1989, wenChiralSpinStates1989, Jiang_TopoSC_2020} and was recently revisited using modern techniques~\cite{Song_DopingChiral_2021,divic_anyonsuperconductivity_2024}. Approaching the QCP from the FCI side, we find an equally natural route to superconductivity: electric charge enters the system in form of vortices $\tilde{\varphi}$, which are soft excitations close to criticality. These vortices naturally fuse to a charge-2 boson (which is the Cooper pair) as the cheapest local excitation, thereby providing a mechanism for anyon superconductivity. We show that when doped to finite density, vortices themselves form quantum Hall states, which demonstrates in detail how superconductors without residual topological order appear in proximity to the quantum critical point. These conventional superconductors can emerge naturally from both the FCI and the SX phase and  (with the notable exception of $\nu=2/3$) are found to be distinct from more exotic superconductors that have been recently proposed~\cite{Shi2024}. 

For simplicity we focus on doping of the $(\bar{1}\bar{1}2)$ and $(332)$ FQH state at fillings $\nu=2/3$ and $\nu=2/5$, respectively. In Appendix~\ref{app:generalCharge2SC}, we generalize our arguments to general $(n\pm1,n\pm1,n)$ states and show that there is always a pathway to simple charge-2e superconductivity that breaks translation symmetry.

\subsection{Doping chargon-vortices into the fractional Chern insulator} \label{sec:dopingVortices}

When describing the critical point in Sec.~\ref{sec:Transition}, we assumed that the bosonic parton $b'(x)$ sees finite flux and forms a bosonic FQH state. In the dual vortex description this finite flux seen by $b'(x)$ is equivalent to starting from a finite density of vortices. When studying finite doping around $\nu=2/3$ and $2/5$ arising from $(\bar{1}\bar{1}2)$ and $(332)$ states with $n=2$, we therefore work in the extended parton picture discussed in Sec.~\ref{sec:EffectiveTheoCritPoint} where $e_\sigma= f_\sigma g_1g_2b$. If the extra fermionic partons $g_{1,2}$ form integer quantum Hall states, the boson will experience zero flux. This is the case {when the flux $\Phi_b/\Phi_0=2/(4\pm1) $, see Eq.~\eqref{eq:fluxChargon} for $n=2$, is equally distributed} among the fermionic partons $g_{1,2}$, such that both are at integer filling $\nu_{g_{1,2}}=1$. The couplings of gauge fields to partons and the respective fillings are outlined in Table~\ref{tab:redFermionicPartons}. The boson $b$ now forms a superfluid in the FQH phase and is gapped in the semion crystal.\footnote{One may identify $g_1 g_2b$ as the single boson $b'$ used to describe the FQH state in Sec.~\ref{sec:parton_theory_fci}.} This construction has the advantage that the vortices of $b$ are now at zero density in the FQH state, making it easier to study finite doping compared to the decomposition in Eqn.~\ref{eq:partons_fb}. The vortices for this boson see finite background flux, which can be resolved by introducing the low-energy valley degrees of freedom for the vortices. Now, doping can be easily studied since both fluxes and densities are proportional to the doping.

After moving to the dual vortex description of Eq.~\eqref{eq:extended_parton_lagrangian}, the Lagrangian of the system reads
\begin{equation}
\begin{aligned}
    \mathcal{L} = &\mathcal{L}_{f_\sigma}[a_1] + \mathcal{L}_{g_1}[a_2-a_1]  +  \mathcal{L}_{g_2}[A-a_2 + a] + \mathcal{L}_{\tilde{\varphi}}[\beta]  \\
    -& \frac{1}{2\pi} a\wedge\mathrm{d}\beta,
\end{aligned}
\end{equation}
where the action for the fermionic partons is given in~\eqref{eq:extended_parton_lagrangian_contd} and the last two terms refer to the vortex fields $\tilde\varphi$. We now analyze this theory for specific cases.

\textbf{\textit{Doping the $\nu=2/3$ state.---}}
Consider doping holes so that the density of up and down electrons per unit cell is $n_{e\uparrow} = n_{e\downarrow} = \frac{1}{3}-\frac{\delta}{2}$. Due to our parton construction Eq.~\eqref{eq:partons_fg1g2b}, this implies a change of all the parton densities. To keep the fermionic partons gapped, we assume that the flux per unit cell of $a_{1,2}$ and $a$ adjusts upon doping:
\begin{equation}
    \begin{aligned}
        \frac{1}{2\pi}\nabla\times a_1& = -\frac{1}{3} + \frac{\delta}{2} \\
           \frac{1}{2\pi}\nabla\times a_2 &= \frac{1}{3} -\frac{\delta}{2} \\     
        \frac{1}{2\pi}\nabla\times a &= -\frac{3\delta}{2},
    \end{aligned}
\end{equation}
which keeps all fermionic partons gapped in integer levels for general $\delta$. Since $\mathrm{d}a/2\pi\rightarrow0$ as $\delta\rightarrow0$, the bosons $b$ see excess flux only when the system is doped, ensuring that the density of vortices of the $b$ field is also proportional to $\delta$. The key question is the filling of vortices, i.e. the ratio of their density to the flux they see. This is $\nu^{\mathrm{total}}_\mathrm{vortices}=3/2$ and will be derived below, but let us give a simple physical argument for it first. The flux seen by the vortices is just the added charge hence $-\delta$. Now, for the density of vortices, we are working in the low-energy space where the only degrees of freedom are vortices which carry electric charge $2e/3$. Thus if we add $\delta$ of charge, this implies a vortex density of $(-\delta)/(-2/3\delta) = 3\delta/2$. The ratio of this vortex density to the flux gives us: $\nu^{\mathrm{total}}_\mathrm{vortices}=3/2$. We will see that there are natural bosonic quantum Hall states that occur at this filling, which will lead overall to a superconductor.
Equations of motion provide the following constraints:
\begin{equation}
    \begin{aligned}
        \mathrm{d}a_1+ \mathrm{d}a_2 &= 0 \\
         \mathrm{d}a -  \mathrm{d}a_2 & = 2\pi \; n_b \\
        2 \mathrm{d}a_2 -  \mathrm{d}a_1 & =  \mathrm{d}a,
    \end{aligned}
\end{equation}
which are satisfied for $n_b = 2/3 -\delta$, as fluxes are only counted modulo 1.\footnote{In a background magnetic field $dA/2\pi =1/2$ one sets $da_1 = -1/6 + \delta/2, \; da_2 =1/6 - \delta/2,  da = 1/2 - 3\delta/2, \; dA = 1/2$, which implies that $n_b = (da-da_2)/2\pi = 1/3 -\delta$, while the bosons see a flux of $dA-da = 3\delta/2$.}
This way, the state formed by the partons $f_\sigma,g_1,g_2$ remains unchanged. We can describe the change of flux in the vortex picture. Assuming all the extra charge enters as vortices, they are at the following density:
\begin{equation}
    n_{\tilde{\varphi}} = \frac{1}{2\pi}\nabla\times a =-\frac{3}{2} \delta,
\end{equation}
this can be seen either from the fact that vortices have charge $2/3$, or from the equations of motion derived from $\frac{\delta \mathcal{L}}{\delta \beta}=0$, relating $n_{\tilde{\varphi}}$ with $\mathrm{d}a/2\pi$ in the same way. The vortices see a background magnetic field equal to:
\begin{equation}
   \frac{\mathrm{d}\beta}{2\pi}  = (-\mathrm{d}a_2 + \mathrm{d}a)\frac{1}{2\pi} = -\frac{1}{3} - \delta,
\end{equation}
which is just the filling. The first term $-1/3$ gives us three valleys for the vortices into which we can dope. The second term is the flux that must be accommodated by forming a bosonic quantum Hall state. Note, the vortices are at filling: $\nu^{\mathrm{total}}_\mathrm{vortices}=3/2$. There are two natural choices for vortices at this filling. At $\nu^{\mathrm{total}}_\mathrm{vortices}=3/2$, they could form a Jain state and spontaneously pick a single valley
\begin{equation}
\begin{aligned}
    \nu_\mathrm{Jain} &= \frac{p}{p+1}; \quad p=-3,  \label{eq:vorticesJain}
\end{aligned}
\end{equation}
or alternatively the vortices can form a valley singlet which is described by:
\begin{equation}
\begin{aligned}
    &K_{\text{v,singlet}} = \begin{pmatrix}
        0 & 1 & 1 \\
        1 & 0 & 1 \\
        1 & 1 & 0 
    \end{pmatrix}, \quad \mathbf{q}^\text{v} = \begin{pmatrix}
        1\\
        1\\
        1
    \end{pmatrix},
\end{aligned}
\label{eq:vorticesSinglet}
\end{equation}
leading to the following action for the vortices
\begin{equation}
      \mathcal{L}_{\tilde{\varphi}}[\beta]= -\frac{1}{4\pi}K_{\text{v,singlet}}^{IJ}\gamma_I\wedge\mathrm{d}\gamma_J + \frac{1}{2\pi}\beta \wedge\mathrm{d} \sum_{I=1}^3\gamma_I.
\end{equation}
The K-matrices for these two states are actually identical, although $K_\mathrm{v,singlet}$ makes the valley singlet structure explicit, highlighting the translational invariance of the doped state.

Upon gluing these {field theories} back together with the fermions, both of these options result in the same Lagrangian 
{\begin{equation}
\begin{aligned}
    \mathcal{L}_{\text{doped}}^{2/3} &= \mathcal{L}_{\tilde{\varphi}}[\beta]-\frac{1}{2\pi}\beta \wedge \mathrm{d}a +\mathcal{L}_{\text{FQH}}[a+A] \\
    &=\mathcal{L}_{\text{FQH}}[a+A] +  \frac{1}{4\pi}K_\mathrm{v}^{IJ}  \gamma_I \wedge \mathrm{d}\gamma_J \\
    &+ \frac{1}{2\pi}\beta \wedge \mathrm{d}  \big(\sum_I  q^\text{v}_I \gamma_I -a\big)
\end{aligned}
\end{equation}}
as the K-matrices of the Jain state and the (00011) singlet are essentially identical. After integrating out $a$ the topology of the resulting superconductor is described by
\begin{equation}
    K^{(2/3)}_{\mathrm{doped}} = \begin{pmatrix}
        1&2&-1&-1 &-1\\
        2&1&-1 &-1&-1\\
        -1&-1&0 & 1 & 1\\
        -1&-1&1 & 0  & 1 \\
        -1&-1&1 & 1 & 0
    \end{pmatrix} ,\; \mathbf{q}_s= \begin{pmatrix}
        1\\
        1\\
        0\\
        0\\
        0
    \end{pmatrix},
\end{equation}
for the field vector $ \left ( \alpha_\uparrow,\,\alpha_\downarrow,\,\gamma_1,\,\gamma_2,\gamma_3\right )$ where the first two fields support the 2/3 FQH state, and the last three account for doping.  For this theory we have $\det  K^{(2/3)}_{\mathrm{doped}} =0$ with zero eigenvalue of eigenvector $(1,1,1,1,1)^T$ . The K-matrix can be block-diagonalized~\footnote{consider 
\begin{equation}
    V = \begin{pmatrix}
       \mathrm{Id}_{4\times4} & 0 \\
         \vec{1}& 1 
    \end{pmatrix}
\end{equation}
with $\det V= 1$.} with a unimodular matrix $\tilde{K}^{(2/3)}_{\mathrm{doped}} = VK^{(2/3)}_{\mathrm{doped}}V^T$
which leaves us with the following theory
\begin{equation}
    \mathcal{L}_{\text{doped}}^{2/3} = \frac{2}{2\pi} \alpha\wedge\mathrm{d}A  -\frac{2}{4\pi}A_s \wedge \mathrm{d}A_s -\frac{2}{4\pi} A \wedge \mathrm{d}A,
\label{eq:doped_112_SC}
\end{equation}
where $\alpha$ is the gauge field corresponding to the zero eigenvalue of $K^{3/2}_\text{doped}$. Integrating it out leads to the Meissner effect $dA=0$. Thus, Eq.~\eqref{eq:doped_112_SC} describes a charge-2e superconductor with no residual topological order. Furthermore, the sign of the quantized spin response tells us that this is a `d-id' superconductor, at least at the level of edge modes, with   $c_-=-2$, i.e. two chiral edge modes propagating opposite to the direction of the charge Hall conductance, consistent with~\cite{Shi2024}.

Finally, let us discuss translation symmetry. This depends on the valley structure of the doped vortices. Forming a singlet of the SU(3) valley symmetry automatically preserves the translational symmetry of the system, since the latter is a subgroup of the former. However, spontaneously choosing a valley and subsequently forming a $p=-3$ Jain state breaks translational symmetry, leading to a SC+CDW phase. Both are possible options that can be realized, which depend on the specifics of the microscopic interaction.

\textbf{\textit{Doping the $\nu=2/5$ state.---}}
To describe the $(332)$ state at $\nu=2/5$, we only need to switch the flux experienced by the spinons, such that they each reside in an IQH state at level $+1$. Keeping all the partons in their respective levels corresponds to the following flux assignment:
\begin{equation}
    \begin{aligned}
        \frac{1}{2\pi}\nabla\times a_1& = \frac{1}{5} - \frac{\delta}{2} \\
           \frac{1}{2\pi}\nabla\times a_2 &= \frac{3}{5} -\frac{3\delta}{2} \\     
        \frac{1}{2\pi}\nabla\times a &= -\frac{5\delta}{2},
    \end{aligned}
\end{equation}
which now implies that the vortices see a flux $\mathrm{\mathrm{d}\beta}/2\pi= -\frac{3}{5}-\delta $ and are at density $n_{\tilde{\varphi}} = -5\delta/2 $. At total filling $\nu^\text{total}_{\text{vortices}} = 5/2=2+1/2$ the vortices can pick a valley in which they form a bosonic IQH state with a 1/2 Laughlin state on top. This vortex state is described by
\begin{equation}
\begin{aligned}
    \mathcal{L}_{\tilde{\varphi}}[\beta] &= -\frac{1}{4\pi} K_{\text{v,}2+\frac{1}{2}}^{IJ}\gamma_I\wedge\mathrm{d}\gamma_J + \frac{1}{2\pi}\beta \wedge\mathrm{d} \sum_{I=1}^3\gamma_I ,\\
   & K_{\text{v,}2+\frac{1}{2}} =  \begin{pmatrix}
        0 & 1 & 0\\
        1 & 0  & 0 \\
        0 & 0 & 2
    \end{pmatrix}, \qquad \mathbf{q}^\text{v} = \begin{pmatrix}
        1 \\
        1  \\
        1 
    \end{pmatrix}.
\end{aligned}
\end{equation}
When combined with $\mathcal{L}_{\text{FQH}}[a]$ this produces a charge-2e superconductor without residual topological order. As in the case of $2/3$ filling, the superconductor inherits the spin edge modes from the spin sector (or equivalently $c_-=+2$), which are now propagating in the {\em same} direction as the charge modes of the FQH state. This pattern persists for all singlet FCIs: conventional charge-2e superconductivity can form if it also breaks translations. Appendix~\ref{app:generalCharge2SC} explains this construction in detail and also presents new families of SC$^*$ states that may arise from the doped FCIs. These translation symmetry-breaking superconductors differ from those proposed in~\cite{Shi2024} at related fillings.

\begin{figure}
\begin{center}
\includegraphics[width=0.99\linewidth]{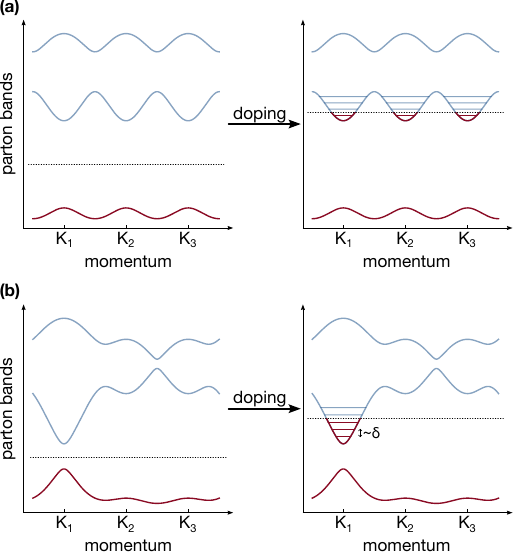}
\caption{\textbf{Doping fermionic partons.} 
Due to the projective action of translations on the partons, there are three degenerate valleys $\mathbf{K}_i$ at $\nu=2/3$. Upon doping, partons in each valley feel a flux $\Delta\Phi \propto \delta$, leading to the formation of Landau levels. \textbf{(a)} Valley-symmetric doping, preserving translations. \textbf{(b)} When translations are broken, either explicitly or spontaneously upon doping, all doped partons enter into a unique valley. 
}
\label{fig:dopingPartons}
\end{center}
\end{figure}

\subsection{Equivalent picture using fermionic partons} \label{sec:dopingPartons}

Our parton construction in the previous Eq.~\eqref{eq:partons_fg1g2b} was chosen to be able to conveniently describe the FQH and semion crystal phases at different fillings, as well as the superconductors that emerge on doping them. Here, we outline how some of the same results at particular fillings can be recovered using just the fermionic partons and dispensing with the boson. This specific parton ansatz was used in~\cite{Shi2024} to analyze the doped 2/3 fractional Chern insulator; here we adapt it to our setting and supply an independent derivation highlighting how a charge-2e superconductor emerges upon doping, generalizing the approach to generic spin singlet FQH states at filling $\nu=2/(2n\pm1)$; see Appendix~\ref{app:generalCharge2SC}. 

Before delving into finite doping, let us briefly specialize to $\nu=2/3$ and discuss how all three phases, FQH, semion crystal and superconductor can be accessed at fixed density within the fermionic parton theory $e_\sigma(x) = f_\sigma g_1g_2$ . While energetics may make a superconductor at the same filling unlikely, an effective theory will help us build intuition for the parton description. Indeed, starting from the parton Chern number assignments $(-2,1,1)$ for the $(f_\sigma,\,g_1,\,g_2)$ partons, it is easily seen that the semion crystal is accessed with the assignments $(-2,1,0)$ while the superconductor with $(-2,1,-2)$. Since this only involves changing the topology of the last parton, that can be modeled as sign changes of Dirac fermion mass, which models the topology-changing transitions. Moreover, the magnetic translation symmetry resulting from the fact that $g_2$ sees emergent gauge flux of $1/3$, implies three valleys related by translations. This gives rise to three species of low-energy fermions $\Psi_i$, which are related to one another by magnetic translations. Ultimately, we can write down the following effective theory:
\begin{eqnarray}
    {\mathcal L}_{\rm Dirac} &=& \sum_{i=1}^3 \bar{\Psi}_i \fsl{\mathcal D}_{a+A} \Psi_i +m_i\bar{\Psi}_i\Psi_i \\ \nonumber &-&\left \{\frac{1}{8\pi}(a+A)\wedge\mathrm{d}(a+A)+CS_g \right \} 
    +\frac{2}{4\pi} a\wedge\mathrm{d}a,
    \label{eq:fixed-densityL}
\end{eqnarray}
where $\fsl{\mathcal D}_a = \gamma\cdot(\partial +ia)$ is the covariant derivative contracted with the Dirac gamma matrices, and the last term arises by noticing that Chern assignments of the first two partons are always $(f_\sigma,\,g_1)= (-2,1)$ implying that they are in the bosonic integer quantum Hall phase with Hall conductance 2~\cite{grover2013}. Finally, the terms in curly brackets (including the gravitational Chern-Simons term $CS_g$, which accounts for chiral edge modes in units of $c_-=1/2$) are introduced to properly offset the topological action, and are at a fractional level since we have an odd number of Dirac fermions. This analysis parallels that of Ref.~\cite{Lee2018}, which studied composite fermion Dirac transitions between FCIs. If all the mass terms are equal and positive, translations are preserved and we obtain the FQH: ${\mathcal L}_{FQH}=\frac{1}{4\pi}(a+A)\wedge\mathrm{d}(a+A)+CS_g+\frac{2}{4\pi} a\wedge \mathrm{d}a$ which has Hall conductance 2/3 and no edge central charge~\cite{Shi2024}. On the other hand, if all mass terms are equal and negative, again we have translation symmetry, but now the action is: ${\mathcal L}_{SC}=-\frac{2}{4\pi}(a+A)\wedge\mathrm{d}(a+A)-4CS_g+\frac{2}{4\pi} a\wedge\mathrm{d}a$, which is readily seen to be a 2e-superconductor with $c_-=-2$. Finally,  we consider the semion crystal. The Chern number assignment requires that we make one mass term negative and the other two positive. This choice however breaks translation symmetry, and if it occurs spontaneously, corresponds to crystallization. Unlike the critical bosonic QED$_3$ theory Eq.~\eqref{eq:L_QCP_effective}, where topology change and translation symmetry breaking occur together, in these variables that scenario is less natural. The effective theory then is: ${\mathcal L}_{\text{SX}}=\frac{2}{4\pi} a\wedge \mathrm{d}a$. It is worth noting two points. First, at the fixed density transition from FQH to superconductor, we will have enhanced correlations at the wave-vectors connecting the three valleys, resulting in an `octet' of CDW correlations corresponding to  (traceless part of) the fermion bilinears $\langle \bar{\Psi}_i\Psi_j\rangle$\cite{Lee2018}. This is despite the fact that at this transition the actual CDW phase is not in the picture. Second, the $2/3$ insulator is special in having such a remarkably simple description of all three phases, and even a fourth one. For the latter, two Dirac masses are negative and one positive, and the system enters an integer quantum Hall insulator that breaks translation and exhibits $\sigma_{xy}=-2$. This simplicity follows from the mapping in Section \ref{sec:EffecTheoryCooperPairs} that relates all these phases to quantum Hall states of Cooper pairs. Furthermore, all of them correspond to Jain states for the Cooper pairs, and hence can be accessed by tuning the Chern number of the associated composite fermions. Indeed, the $\Psi$ are the composite fermions obtained after attaching flux to the Cooper pairs. Now, let us turn to the finite doping transition, where these Chern assignments will be obtained by adding charge.

 \textbf{\textit{ Density tuned transitions from fermionic partons.---}} For simplicity, we consider particle doping $\delta > 0$. To form a superconductor upon doping, the Chern number of one of the auxiliary partons, say $g_2$, must change from $1$ to $-2$. To keep both the spinons $f_\sigma$, and the other chargon $g_1$ at a fixed integer levels $(-2,1)$ requires the emergent gauge fields to adjust such that $g_2$ experiences a magnetic flux of $n_{\Phi}^{g_2} = \Phi_{g_2}/\Phi_0  = -\frac{1}{3}-\frac{\delta}{2}$, while being at a density of $n_{g_2} = \frac{2}{3}+\delta$. In the presence of a lattice, the total Chern number of $g_2$ is constrained by a Diophantine equation~\cite{TKNN, Dana_1985, agazziColoredHofstadterButterfly2014}:
\begin{equation}
    C_{g_2} \cdot\;n_{\Phi_{g_2}} -n_{g_2} = \mathbb{Z}.
    \label{eq:diophantine}
\end{equation}
For Eq.~\eqref{eq:diophantine} to be satisfied, terms involving the doping $\delta$ must cancel while also recovering $C_{g_2}=1$ in the limit $\delta\rightarrow 0$. Thus, we find $C_{g_2}=-2$ as a consistent solution in the doped case (see Appendix~\ref{app:diophantine} for more details), which signals the formation of charge-2e superconductivity. This illustrates, why the underlying lattice potential is crucial for establishing superconductivity upon doping the FCI. 
For the FCI at $2/3$ filling, this establishes a relation to previous work~\cite{Shi2024}, which we obtain in an independent approach. 
 Performing a continuum expansion and taking into account that due to the flux seen by $g_{1,2}$, there are three equivalent minima $\mathbf{K}_i$, we find
\begin{equation}
    H_{\mathbf{K}} = \sum_{i=1}^3 \frac{1}{2 m^*}(\mathbf{k} - \mathbf{K}_i)^2 + \cdots \label{eq:dopedvalley_Hamiltonian}
\end{equation}
Upon doping, the finite spin gap close to the critical point ensures that the spinons remain in their gapped IQH state. Consequently, it is energetically favorable to dope spin-singlet anyons which carry charge $2/3$ and correspond to $g_{1,2}$. When changing the electron density by $\delta$, we need to adjust the densities of all partons $f_\sigma,g_1,g_2$. To ensure that the spinons remain in a gapped level $-2$ IQH state, we must therefore change the flux they experience. Specifically, their flux changes by $- \delta/2$ according to the Streda formula. Since the partons $g_{1,2}$ are coupled to the spinons through the gauge fields $a_{1,2}$,
\begin{equation}
    \mathcal{L} = \mathcal{L}_{f_\sigma}[a_1] +  \mathcal{L}_{g_1}[a_2-a_1] +  \mathcal{L}_{g_2}[-a_2+A],
\end{equation}
a change in the flux of the spinons leads to a change in the flux of the partons $g_{1}$ and $g_2$.
As discussed, the flux of the spinons must change by 
\begin{equation}
   \frac{1}{2\pi} \nabla \times a_1 = -\frac{1}{3}- \frac{\delta}{2}
\label{eq:spinons_doped_flux}
\end{equation}
such that they remain in an IQH state. 
We show that choosing a change in flux for the gauge field
\begin{equation}
   \frac{1}{2\pi} \nabla \times a_2 = \frac{1}{3}+ \frac{\delta}{2}
\label{eq:specific_doped_flux}
\end{equation}
results in the same superconductor as in Sec.~\ref{sec:dopingVortices}. Note that the flux assignment Eq.~\eqref{eq:specific_doped_flux} is consistent with no change in the external flux of $A$. The flux $\Phi_{g_2}$ seen by $g_{2}$ changes by $\Delta\Phi_{g_2}/\Phi_0 = -\delta/2$, adjusting the Hamiltonian 
\begin{equation}
    H_{\mathbf{K}} = \sum_{i=1}^3\frac{1}{2 m^*}(\mathbf{k} - \mathbf{K}_i - \mathbf{a}_\delta)^2 + \cdots \label{eq:dopedvalley_Hamiltonian_adjusted}
\end{equation}
where $\mathbf{a}_\delta$ is a gauge field generating flux $\Delta \Phi_{g_2}$. Consequently, Landau levels form in the doped valleys, where the cyclotron frequency scales with doping $\Delta\Phi_{g_2} \propto-\delta$. This suggests that the stiffness of the resulting superconductor will scale with doping.
 With this flux assignment for $a_{1/2}$, parton $g_1$ remains in a $C=1$ band after doping, while $g_2$ fills an integer number of Landau levels in the doped valleys, such that its total Chern number is $C=-2$; see Table~\ref{tab:doping_mmn_fermionic} for an overview. 
This can be seen as follows: After doping, the density for $g_{2}$ is  $n_{g_2} = 2/3+\delta$ per unit cell, and it feels a flux of $\Phi_{g_2} = 1/3 - \delta/2$ in units of $\Phi_0$. 
On the FQH side of the transition, the filled band has $C=1$. The Streda formula implies that a change in the flux leads to a change in the density of partons that fill that band. Hence, we find that only $n_{C=1} = 2/3 - \delta/2$ partons fit in the originally filled band, leaving a density of $\Delta n = n_g - n_{C=1} = 3\delta/2$ to fill the valleys in the second band. When the valley symmetry of the partons is preserved, the doped partons fill $\nu=\Delta n/(3\Delta \Phi) = -1$ Landau levels in all three valleys. Alternatively, if translations are broken, either explicitly or spontaneously upon doping, valley symmetry is broken. In this case, all doped partons enter into a single valley, where they fill $\nu = \Delta n/\Delta \Phi = -3$ Landau levels; see Fig.~\ref{fig:dopingPartons}. Doping in a valley-broken state corresponds to the vortices forming a Jain state, Eq.~\eqref{eq:vorticesJain}, while doping the partons in a valley-symmetric state corresponds to the vortices forming a valley singlet, Eq.~\eqref{eq:vorticesSinglet}. 

On the semion crystal side of the transition, the filled band for $g_2$ has $C=0$, such that all doped partons $\Delta n = \delta$ enter into the unique minimum of the second band, Eq.~\eqref{eq:dopedvalley_Hamiltonian}, where they fill $\nu = \Delta n/\Delta \Phi = - 2$ Landau levels.

The final Chern number for the $g_2$ partons is the sum of the originally filled lower Chern band and the Landau levels in the doped valley, giving $C=-2$ in total, on both sides of the transition, as suggested by the Diophantine equation Eq.~\eqref{eq:diophantine}. 
This highlights the importance of a lattice or at least a periodic potential, as the same arguments cannot be applied to a system with continuous translations in flat Landau levels. 

Putting everything together, the resulting Lagrangian after doping is
\begin{equation}
\begin{aligned}
        \mathcal{L} =&  \frac{1}{4\pi}\Big[\alpha_\uparrow \wedge \mathrm{d}\alpha_\uparrow
    + \alpha_\downarrow \wedge \mathrm{d}\alpha_\downarrow -\gamma \wedge \mathrm{d}\gamma  \\
    & \qquad + \beta_1 \wedge \mathrm{d}\beta_1
    + \beta_2 \wedge \mathrm{d}\beta_2 \Big] \\
    &+\frac{1}{2\pi} \Big[ a_1 \wedge \mathrm{d}(\alpha_\uparrow + \alpha_\downarrow -\gamma) \\
    & \qquad +  a_2 \wedge \mathrm{d}(\gamma - \beta_1- \beta_2)  + (\beta_1 + \beta_2)\wedge \mathrm{d}A \Big].
\end{aligned}
\end{equation}

Integrating out the gauge fields $a_{1/2}$ enforces the constraints $\gamma = \beta_1 + \beta_2 = \alpha_\uparrow + \alpha_\downarrow$. Defining $\beta \equiv\beta_1$, we find 
\begin{equation}
    \begin{aligned}
        \mathcal{L}^{2/3}_\text{doped} &=  \frac{2}{4\pi} \beta \wedge \mathrm{d}\beta + \frac{1}{2\pi} (\alpha_\uparrow+\alpha_\downarrow) \wedge \mathrm{d}\beta \\
        &+ \frac{1}{4\pi} (\alpha_\uparrow\wedge \mathrm{d}\alpha_\uparrow + \alpha_\downarrow\wedge \mathrm{d}\alpha_\downarrow) + \frac{1}{2\pi}(\alpha_\uparrow+\alpha_\downarrow) \wedge \mathrm{d}A
    \end{aligned} \label{eq:LagrangianDoped}
\end{equation}
which has a K-matrix and charge vector of
\begin{equation}
   K_\mathrm{doped}=  \begin{pmatrix}
       - 1 & 0 & -1\\
        0 & -1 & -1\\
        -1 & -1& -2
    \end{pmatrix}, \; \mathbf{q} =\begin{pmatrix}
       1\\
        1 \\
        0 
    \end{pmatrix}
\end{equation}
for which $\det K = 0$. The matrix can be diagonalized with a unimodular matrix $\tilde{K} = WKW^T$,\footnote{
\begin{equation}
 W = \begin{pmatrix}
       1 & 1 & -1\\
        0 & 1 & 0\\
        1 & 0& 0
    \end{pmatrix} \quad \text{with} \quad \det W =1 
\end{equation}

}

which leaves us with the same theory, Eq.~\eqref{eq:doped_112_SC}, as in Sec.~\ref{sec:dopingVortices}, describing a charge-2e superconductor, without residual topological order and a spin quantum Hall response of $-2$.
Finally, we argue that the doped state has zero electrical resistivity, as required for a superconductor. To that end, note that gluing the partons $g_1$ and $g_2$ at level (1,-2) back together after doping, can be viewed as forming a bosonic IQH state for the composite boson $g_1g_2$, 
characterized by a Hall conductivity of $\sigma_{xy}=+2$~\cite{Lu_TheoryClassification_2012, Senthil_bIQH_2013, grover2013}. This exactly cancels the Hall resistivity of the spinons, resulting in a vanishing total resistivity according to the Ioffe-Larkin rule: $\rho_e = \rho_b + \rho_f = 0$~\cite{Song_DopingChiral_2021}.

\begin{table}[]
    \centering
        \begin{tabular}{c|c|c|c}
        \toprule
        species & density &  flux  & filling \\
         & (w.r.t. lattice) & (per unit cell) & \\
        \hline
        $f_\sigma$ & $\frac{1}{6}+\frac{\delta}{2}$  & $-(\frac{1}{6} +\frac{\delta}{2})\Phi_0$ & $-2\rightarrow-2$ \\
        % \hline
        $g_1$ & $\frac{1}{3}+\delta$ & $(\frac{1}{3} +\delta)\Phi_0$ & $ +1\rightarrow +1$\\
        % \hline
        $g_2$ & $\frac{1}{3}+\delta$& $(\frac{1}{3} -\frac{\delta}{2})\Phi_0$ & $+1\rightarrow-2$ \\
        \bottomrule
        \end{tabular}     
    \caption{Doping fermionic partons in the $(\bar{1}\bar{1}2)$ state at $\nu=2/3$. The flux is adjusted upon doping, such that the spinons remain in their original IQH state. The total Chern number (filling) before and after doping is shown. For simplicity, we assume a constant background flux of $\Phi_0/2$ per unit cell.}
    \label{tab:doping_mmn_fermionic}
\end{table}

\textbf{\textit{Doping the $\nu=2/5$ state.---}}
In the $(332)$ state, the spinons have total Chern number $C=+2$. Applying similar reasoning as for the $(\bar{1}\bar{1}2)$ state, adjusting the flux for the spinons such that they remain in their gapped $C=2$ state, introduces a change in flux for the partons $g_{1/2}$, as outlined in Table~\ref{tab:doping_332_fermionic}. Assuming that all doped partons enter into a unique valley, which implies translation symmetry breaking we find that $g_1$ has total Chern number $C=-1$ after doping, and $g_2$ has $C=2$. The resulting superconductor is the same as for $\nu=2/3$, Eq.~\eqref{eq:doped_112_SC}, but with a spin Hall response of $+2$ instead of $-2$. 
While the superconductor, with broken valley symmetry for the partons, corresponding to broken translations for the electronic state, straightforwardly generalizes from $\nu=2/3$ to $\nu=2/5$, the {\em valley-symmetric state} of $\nu=2/3$ does not have a counterpart at $\nu=2/5$, where such valley symmetric states map generically to more exotic superconductors. From this perspective, the $\nu=2/3$ state is special, as here the valley-symmetric and valley-broken superconductors are topologically equivalent.

\begin{table}[]
    \centering
        \begin{tabular}{c|c|c|c}
        \toprule
        species & density &  flux  & filling \\
         & (w.r.t. lattice) & (per unit cell) & \\
        \hline
        $f_\sigma$ & $\frac{1}{10}+\frac{\delta}{2}$  & $(\frac{1}{10} +\frac{\delta}{2})\Phi_0$ & $+2\rightarrow+2$ \\
        % \hline
        $g_1$ & $\frac{1}{5}+\delta$ & $(\frac{1}{5} -\delta)\Phi_0$ & $ +1\rightarrow -1$\\
        % \hline
        $g_2$ & $\frac{1}{5}+\delta$& $(\frac{1}{5} +\frac{\delta}{2})\Phi_0$ & $+1\rightarrow+2$ \\
        \bottomrule
        \end{tabular}     
    \caption{Doping fermionic partons in the $(332)$ state at $\nu=2/5$. The flux is adjusted upon doping, such that the spinons $f_\sigma$ remain in their original IQH state. The total Chern number (filling) before and after doping is shown. For simplicity, we assume a constant background flux of $\Phi_0/2$ per unit cell.}
    \label{tab:doping_332_fermionic}
\end{table}

\subsection{Doping the semion crystal}
Close to the critical point, the semion crystal is described by the following action
\begin{equation}
\mathcal{L}_{\text{SX}}=\mathcal{L}_\varphi[-\alpha+A] \pm \frac{2}{4\pi} \alpha\wedge\mathrm{d}\alpha +\frac{2}{2\pi}A_s \wedge\mathrm{d}\alpha,
\label{eq:dopingSemions}
\end{equation}
where we have explicitly reinstated a charge $-e$ bosonic field $\varphi$, corresponding to the chargon in the parton decomposition $e_\sigma = f_\sigma b'$, which is in a translation symmetry-breaking Mott phase but close to its critical point. {Doping this chiral spin-liquid works similarly to previous proposals~\cite{Song_DopingChiral_2021,divic_anyonsuperconductivity_2024}, with the difference of translational symmetry breaking in the semion crystal we discuss.} In a regular Kalmeyer-Laughlin spin liquid, the charge gap is much larger than the spin gap $\Delta_c\gg\Delta_s$, which implies that the lowest lying anyons carry spin 1/2. In stark contrast, the lowest energy excitations of the semion crystal close to the QCP are spinless, charge $Q=-e$ semions described by $\varphi(x)$. Since $\Delta_c/\Delta_s \rightarrow 0$ it is natural to think of these anyons as fractionalized Cooper pairs. As Sec.~\ref{sec:EffecTheoryCooperPairs} shows, a similar interpretation is possible on the FCI side of the transition.

Upon hole doping, the density of doped chargons is $-\delta$. To keep the spinons gapped, we must adjust the flux $\frac{1}{2\pi}\nabla\times \alpha \sim \mp \delta/2$, leading to a filling $\nu_b = \pm 2$. Consequently, the $b$ bosons form a bosonic integer quantum Hall state, which when integrated out precisely cancels the topological order in~\eqref{eq:dopingSemions} arising from the spinons. We expect that the translational symmetry breaking is preserved upon doping, which leads to a charge-2e superconductor without residual topological order in the presence of a charge density wave, see Fig.~\ref{fig:overview}(b).

Forming any bosonic state with $\sigma_{xy}^b=-2$ will give rise to a superconducting state according to the Ioffe-Larkin rule. Which state is chosen then influences the residual topological order. If the bosons were to form a more exotic $\nu_b=-2$ state, such as a Laughlin state of paired semions, the resulting state is a SC$^*$ as it retains background semion topological order.

\subsection{Properties of the anyon superconductor}
At $2/3$ filling, both options we discussed are topologically `$d-id$' superconductors. Owing to the fact that the neutral spin modes on the boundary are counter-propagating and remain unmodified close to the critical point, the superconducting state retains a spin response $-\frac{2}{4\pi}A_s \wedge \mathrm{d} A_s$ which is opposite to the electronic Hall effect. Due to the nature of the FQH parent states at general fillings $\nu = 2/(2l+1)$, the sign of the spin-quantum Hall effect (and therefore also the thermal Hall response) alternates as $\sigma_{xy}^s= 2\cdot (-1)^l$. {Interestingly, while the spin-quantum Hall response is inherited from the parent states, the chiral central charge changes, i.e. from $c_-=0$ in the $2/3$ FQH state to $c_-=-2$ in the superconductor.}

Due to the finite spin gap, the decay of superconducting correlations is constrained by a finite spin correlation length $\xi_s$:
\begin{equation}
    \langle \Delta(\mathbf{r})\rangle = \langle e_\sigma(0) e_{\bar{\sigma}}( \mathbf{r})\rangle \sim \exp(- \frac{|\mathbf{r}|}{\xi_s}).
\end{equation}
As we are describing an insulator-to-insulator transition, this behavior is expected. Remarkably, since our doping procedure preserves the spin-gap, we expect that it also sets the size of Cooper pairs for weak doping. {This implies Cooper pairs are inherently tightly bound close to the transition, while the topology is consistent with a chiral $d+id$ superconductor in the BCS limit. The tight binding of the Cooper pairs implies that the critical temperature $T_c$ is set by the stiffness $\rho_s$ of the superconductor, rather than the pairing-scale. Starting from each parent state, the stiffness is expected to grow with the doping level and suppressed by the anyon mass $m_*$ such that $\rho_s \sim \delta,\; m_*^{-1}$, see Ref.~\cite{chenAnyonSuperconductivity1989} for a derivation.}
%This implies that the superconductors we describe are inherently in a Bose-Einstein Condensate limit and the critical temperature is set by the stiffness which grows with the doping level.

Which type of superconducting state is selected upon doping, i.e. whether it breaks translations or not, is ultimately a matter of microscopics.

Furthermore, we note that charge-2e superconductivity with opposing thermal Hall response can be obtained by considering Kohn-Luttinger mechanisms in the complementary weakly interacting limit where the parent state is a chiral metal~\cite{maymann_2025,Shavit_2025}.

\section{Microscopic lattice model} \label{sec:Numerics}

 \begin{figure}
\begin{center}
\includegraphics[width=0.99\linewidth]{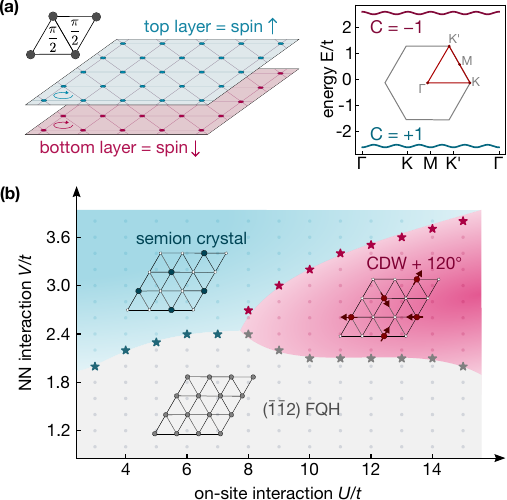}
\caption{\textbf{Model and phase diagram.} \textbf{(a)} Left: Bilayer Hofstadter model on a triangular lattice with $\pi/2$ flux per triangle. The layer degree of freedom serves as a pseudo-spin, which is a good quantum number because we consider vanishing inter-layer hopping. Right: Flat Chern bands of the model for $t'=t/4$. The cut through the Brillouin zone is shown as an inset. \textbf{(b)} Phase diagram for $\nu=2/3$ as a function of the repulsive on-site interaction $U$ and the nearest-neighbor interaction $V$, obtained from Tensor Network simulations on an infinite cylinder with $L_y=3$ and $t'=t/4$, bond dimension $\chi=1024$. For small $V$, we find a stable $(\bar{1}\bar{1}2)$ FQH phase. Increasing $V$ leads to the formation of a CDW, whose spin order depends on $U$. For large $U$, the spins order to form a $120$ degree state. For smaller $U$, the spin-sector forms a CSL {on top of the CDW, i.e., the semion crystal}.}
\label{fig:model}
\end{center}
\end{figure}

\begin{figure}
\begin{center}
\includegraphics[width=0.99\linewidth]{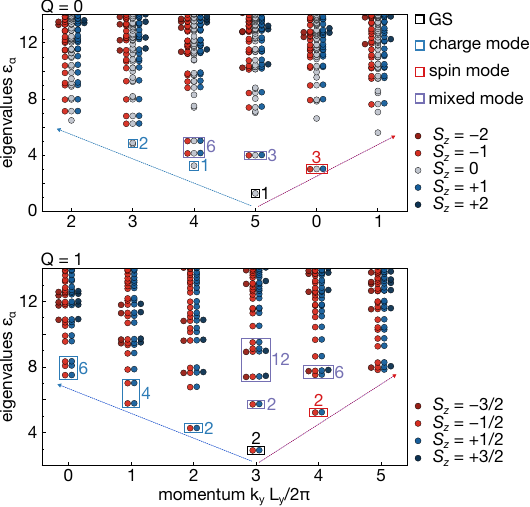}
\caption{\textbf{Entanglement spectrum of the $(\bar{1}\bar{1}2)$-state.} The two chiral edge modes of the $(\bar{1}\bar{1}2)$-state have opposite chirality. Top (bottom) panel shows the charge $Q = 0$ $(1)$ sector. The counting of the lowest states agrees well with {the edge theory. Here, we used} $U=1.2t$, $V=U/2$, $\chi=10000$, $L_y=6$.}
\label{fig:EntanglementSpectrum}
\end{center}
\end{figure}

\begin{figure*}
\begin{center}
\includegraphics[width=0.99\linewidth]{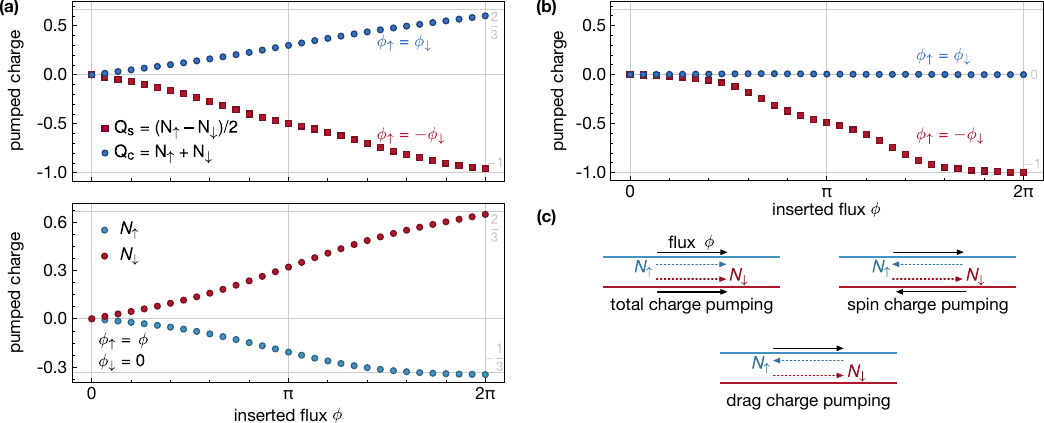}
\caption{\textbf{Charge pumping.} \textbf{(a)} Various charge pumping schemes in the $(\bar{1}\bar{1}2)$ FQH phase, for $U=4t$, $V=0.8t$. Top panel: total charge $Q_c$ and spin charge $Q_s$ pumped after inserting one flux quantum through the cylinder. Bottom panel: Drag charge pumping. \textbf{(b)} Charge pumping in the semion crystal phase, for $U=4t$, $V=4t$. The total charge is frozen, while the spin pumping is the same as for the $(\bar{1}\bar{1}2)$ FQH phase, reflecting that the spin sector does not change across the transition. \textbf{(c)} Illustration of different charge pumping schemes. For total charge pumping, the same flux $\phi$ is inserted in both layers, while for spin pumping, the opposite flux is inserted in the two layers $\phi_\uparrow=-\phi_\downarrow$. Calculations were performed on an infinite cylinder with $L_y=4$ and bond dimension $\chi=1024$.}
\label{fig:cpump}
\end{center}
\end{figure*}

Guided by these general insights, we aim to design a microscopic model that hosts a quantum phase transition from a Halperin FQH state to a semion crystal. To this end, we consider a triangular lattice Hubbard-Hofstadter model of pseudo spins that couple uniformly to a perpendicular magnetic field and form Chern bands. The choice of the triangular lattice is motivated by the fact that the resulting CDW states frustrate spin interactions and moreover are naturally realized in moir\'e lattices of semiconductor heterostructures. By tuning inter- and intralayer interactions, distinct competing phases may be stabilized, such as the Halperin FQH states, CDW states, and semion crystals. With this, we arrive at the following Hamiltonian~\cite{Kuhlenkamp24}:
\begin{equation}
\begin{aligned}
\hat{H} &= - t \sum_{\mathclap{\langle ij\rangle, \sigma = \lbrace \uparrow,\downarrow \rbrace}} e^{i \phi_{ij}} \; c^\dagger_{i,\sigma} c_{j,\sigma} - t' \sum_{\mathclap{\langle\langle ij\rangle\rangle, \sigma = \lbrace \uparrow,\downarrow \rbrace}} e^{i \phi'_{ij}} \; c^\dagger_{i,\sigma} c_{j,\sigma} + \mathrm{h.c.}\\
&+ U \sum_{i,\sigma \neq \sigma'} n_{i,\sigma} \; n_{i,\sigma'}+ V \sum_{\langle ij\rangle,\sigma  \sigma'} n_{i,\sigma} \; n_{j,\sigma'},
\end{aligned}
\label{eq:microscopic-model}
\end{equation}
where $t$ ($t'$) is the (next-)nearest neighbor hopping, $U$ are interlayer interactions and $V$ nearest-neighbor interactions both within and between the layers.
 The Peierls phases are chosen to match the flux through a triangle $\Phi = B_z a^2 \sqrt{3}/4$, where $a$ is the lattice constant. At $\Phi = \frac{\Phi_0}{2}$ flux piercing a triangle, and $ t' = t/4$ the system exhibits nearly flat particle-hole symmetric valence and conduction bands, with Chern numbers $C=1$ and $C=-1$, respectively; see Fig.~\ref{fig:model}(a). Our model explicitly breaks time-reversal symmetry, and the Chern numbers for the two flavors $\sigma$ are equal. At finite filling and intermediate interaction strength, the system can thus realize various bilayer quantum Hall states. The Hubbard-Hofstadter model discussed here should be thought of as a controlled system, with potential experimental implementation in moir\'e heterostructures~\cite{Kuhlenkamp24}. It serves as proof of principle that the considerations of the previous sections can be realized in a microscopic lattice model, but we expect our mechanism to be applicable more generally.

We represent the many-body wave function of the system using matrix product states (MPS) on an infinite cylinder geometry. Using the TeNPy library~\cite{tenpy}, we perform infinite Density Matrix Renormalization Group (iDMRG) calculations to variationally optimize the MPS. 
There exists extensive numerical evidence for a continuous transition in the simple $n=0$ case at a total filling of $\nu=2$~\cite{Kuhlenkamp24, Divic24, divic_anyonsuperconductivity_2024}. Here, we focus on a filling of $\nu=2/3$, corresponding to an electron density of $n_e = \nu \Phi/\Phi_0 = 1/3$ per unit cell. At this filling, a transition from the $(\bar{1}\bar{1}2)$ bilayer FQH state to a semion crystal can be observed. For a small nearest-neighbor interaction $V$ and finite on-site interaction $U$, the $(\bar{1}\bar{1}2)$ FQH state is stabilized. Increasing $V$ favors the formation of a charge-ordered CDW state; see Fig.~\ref{fig:model}(b) for an overview of the phase diagram. The spin order of the CDW depends on the interaction $U$. For large $U$, the dominant spin coupling is the anti-ferromagnetic Heisenberg coupling $J_H \sim \mathcal{O}(t'^2/U, t^4/(V^2 U))$, leading to a spin-ordered $120$ degree phase. Here, the spin exchanges arise on top of the CDW. For example, the `direct exchange' term is realized either by next-nearest neighbor hopping processes or by fourth-order nearest neighbor hopping processes. For smaller $U$, the presence of a chiral coupling $J_\chi \sim \mathcal{O}( t'^3/U^2, t't^4/(V^2 U^2))$, which is suppressed by an additional factor of {$t'/U$} relative to $J_H$, favors the formation of a chiral spin liquid on top of the CDW, {i.e., the} semion crystal.

While the transition between the $(\bar{1}\bar{1}2)$ FQH state and the $120$ degree N\'eel state appears to be first order, we provide evidence for a continuous transition from the $(\bar{1}\bar{1}2)$ FQH state to the semion crystal, with the spin correlation length remaining finite across the transition.

First, we establish that the Halperin $(\bar{1}\bar{1}2)$ state is stabilized in the bilayer Hubbard-Hofstadter model Eq.~\eqref{eq:microscopic-model} on the triangular lattice. In contrast to the $(110)$ state discussed in Refs.~\cite{Kuhlenkamp24, divic_anyonsuperconductivity_2024}, the $(\bar{1}\bar{1}2)$ state has a non-trivial topological order and has fascinating properties in its own right. As discussed in Sec.~\ref{sec:parton_theory_fci}, it can be understood in a parton language as spinons filling Chern bands with $C=-1$ and chargons in a bosonic $1/2$ Laughlin state. 
We uniquely identify the $(\bar{1}\bar{1}2)$ state from its entanglement spectrum and by performing adiabatic flux insertion into the cylinder. 

The entanglement spectrum is characterized by the edge Lagrangian 
\begin{equation}
    \mathcal{L}_\text{edge} = -\frac{1}{4\pi}( K_{IJ} \partial_t \varphi_I \partial_y \varphi_J + V_{IJ} \partial_y \varphi_I \partial_y \varphi_J) \label{eq:edgelagrangian}
\end{equation}
where $V_{IJ}$ is a non-universal positive definite velocity matrix~\cite{wenTopologicalOrdersEdge1995a}. For singlet FQH states, the two edge modes separate into a charge and spin mode, which can be seen by introducing the basis change $\varphi_{c/s} = (\varphi_\uparrow \pm \varphi_\downarrow)/\sqrt{2}$, diagonalizing the edge Lagrangian~\cite{furukawa13, He15}:
\begin{equation}
    \begin{aligned}
        \mathcal{L}_\text{edge} = -\frac{1}{4\pi}[&(2n\pm1) \partial_t \varphi_c \partial_y \varphi_c + v_c \partial_y \varphi_c \partial_y \varphi_c \\
        &\pm \partial_t \varphi_s \partial_y \varphi_s + v_s \partial_y \varphi_s \partial_y \varphi_s] \label{eq:edgeLagrangian}
    \end{aligned}
\end{equation}
for $(mmn)$ states with $m=n\pm1$ and velocities $v_{c/s}>0$. Note that for $m=n+1$, the two edge modes have the same chirality, while for $m=n-1$, they have opposite chirality, resulting in a non-chiral entanglement spectrum, assuming $n>0$. We find the edge Hamiltonian and total momentum~\cite{furukawa13}
\begin{equation}
    \begin{aligned}
        &\mathcal{H}_\text{edge} = \frac{2\pi}{L_y} \bigg(\frac{v_c}{2n\pm 1} L_c + v_s L_s\bigg) \\
        &\mathcal{P} = -\frac{2\pi}{L_y}(L_c \pm L_s) \label{eq:edgeHamAndMomentum}
    \end{aligned}
\end{equation}
with 
\begin{equation}
    L_c = \frac{Q^2}{4} + \displaystyle\sum_{\ell=1}^\infty \ell n_{c, \ell} \quad \text{and} \quad
    L_s = S_z^2 + \displaystyle\sum_{\ell=1}^\infty \ell n_{s, \ell}.
\end{equation}
Here $n_{c/s, \ell} = a^\dagger_{c/s, \ell}a_{c/s, \ell}$ is the occupation of the $\ell$-th charge/spin mode, $Q=N_\uparrow+N_\downarrow$ is the total charge, $S_z = (N_\uparrow-N_\downarrow)/2$ the total spin and $L_y$ the length of the edge.

Using the above, the edge spectrum accessible through the entanglement spectrum can be understood, see Appendix~\ref{app:CountingEdgeSpectrum} for details. There are three dispersing modes: A pure charge mode with $L_s=0$, a pure spin mode with $L_c=0$, and a mixed mode with $L_c\neq 0 \neq L_s$. The counting of the entanglement spectrum in Fig.~\ref{fig:EntanglementSpectrum} agrees well with the expected edge spectrum for the $(\bar{1}\bar{1}2)$ state for which $m=1,\,n=2$, as graphically illustrated.

Generically, $(mmn)$ Halperin states with $n\neq 0$ are interlayer correlated, implying that adiabatically inserting flux in one layer leads to charge pumping in both layers~\cite{zengTwocomponentQuantumHall2017, zengHalperinFractionalQuantum2023}. Analogously, in transport experiments, driving a Hall current in one layer leads to a finite drag-Hall response also in the other layer. Measuring a quantized drag-Hall current numerically by adiabatic flux insertion and in experiments with drag-transport measurements is consequently a powerful tool to detect Halperin states. 

We consider three different charge pumping procedures and simulate them numerically using tensor network methods~\cite{Zaletel2014}. 

(i)
    Total charge pumping: The same flux $\phi = \phi_\uparrow = \phi_\downarrow$ is inserted adiabatically in both layers. After inserting one flux quantum, a total charge 
    \begin{equation}
        Q_c = N_\uparrow + N_\downarrow = \mathbf{q}^T_c K^{-1} \mathbf{q}_c = \frac{2}{m+n}
    \end{equation}
    is pumped. Here $\mathbf{q}_c^T = (1, 1)$ is the charge vector and $N_\sigma$ is the charge pumped in the individual layers. 

    (ii) Drag charge pumping: A flux $\phi_\uparrow = \phi$ is inserted in one layer, while no flux is inserted in the other layer $\phi_\downarrow = 0$. In a system with interlayer correlations, charges are still pumped in both layers:
    \begin{equation}
        \begin{aligned}
            N_\uparrow &= \mathbf{q}^T_\uparrow K^{-1} \mathbf{q}_\uparrow = \frac{m}{m^2 - n^2} \\
            N_\downarrow &= \mathbf{q}^T_\downarrow K^{-1} \mathbf{q}_\uparrow = -\frac{n}{m^2 - n^2}
        \end{aligned} \label{eq:dragcpump}
    \end{equation}
    with $\mathbf{q}^T_\uparrow = (1, 0)$ and $\mathbf{q}^T_\downarrow = (0, 1)$.

    (iii) Spin charge pumping: Opposite flux is inserted in the two layers: $\phi_\uparrow = -\phi_\downarrow = \phi$. Leading to a pumping of spin charge
    \begin{equation}
        Q_s = \frac{1}{2}(N_\uparrow - N_\downarrow) = \frac{1}{2} \mathbf{q}^T_s K^{-1} \mathbf{q}_s = \frac{1}{m-n}
    \end{equation}
    with spin vector $\mathbf{q}_s = (1, -1)$. For the singlet states with $m = n \pm 1$, the spin charge is quantized to $Q_s = \pm 1$, reflecting the spinons in Chern bands with $C_{f_\sigma}=\pm 1$.
    
Both the numerically observed quantized drag-charge pumping and spin pumping are in perfect agreement with a $(\bar{1}\bar{1}2)$ state; see Fig.~\ref{fig:cpump}(a). Concretely, the total charge pumped is $Q_c = 2/3$ after inserting one flux quantum, while a spin charge $Q_s=-1$ is pumped. When we insert a flux quantum only in the top layer, a charge of $N_\uparrow=-1/3$ is pumped in the top layer and charge $N_\downarrow=2/3$ in the bottom layer, consistent with Eq.~\eqref{eq:dragcpump}. 

Increasing the nearest-neighbor interactions triggers a phase transition and tunes the system into a different insulator, which we will argue is nothing but the semion crystal. While the charge response of this phase is trivial, spin-pumping reveals a quantized spin-Hall response, which is identical to the spin-Hall response of the $(\bar{1}\bar{1}2)$ state; see Fig.~\ref{fig:cpump}(b).
For general semion crystals formed at filling $\nu=2/(2n\pm1)$, the quantum spin Hall response is alternating $Q_s=\pm1$. This can be understood in this limit as lattice translations are broken, and the unit cell is enlarged. Within the enlarged unit cell, the electron and spinon density is integer $(n_e=1)$, while the flux per unit cell of the crystal is $\pm \Phi_0/2$ modulo $2\pi$. Thus, the spinon density in the crystal is now large enough to absorb all of the external flux by forming an IQH state at level $\pm2$, while the chargons are in a Mott state with respect to the enlarged unit cell. As a concrete example, consider the 2/3 FCI, which transitions into the semion crystal, shown in the top left of the phase diagram of Figure~\ref{fig:model}(b). While triangles of the lattice enclose $\pi/2$ flux, the larger triangles defined by the charge ordered state have three times the area and enclose a flux of $3\pi/2 \equiv -\pi/2$, which reverses the sign of the spin Hall conductance compared to the chiral spin liquid at integer filling.

\begin{figure*}
\begin{center}
\includegraphics[width=0.99\linewidth]{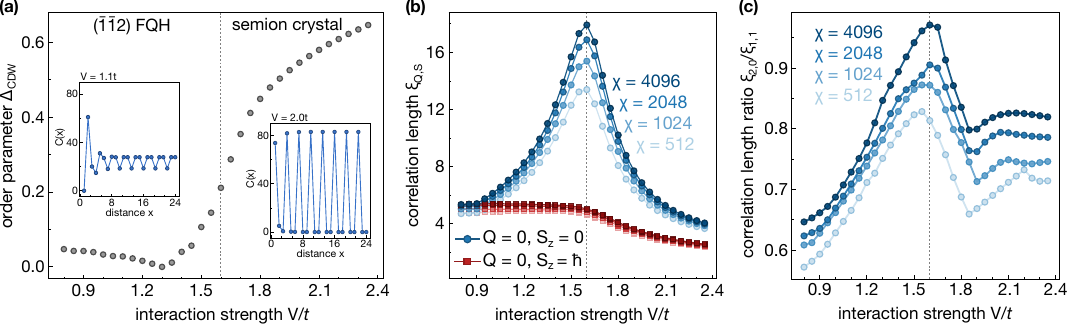}
\caption{\textbf{Evidence for second-order phase transition in the Hubbard-Hofstadter model.}  \textbf{(a)} Order parameter $\Delta_\text{CDW}$ as a function of the interaction strength $V$ for fixed $U=4t$, $\chi=2048$ {and $L_y=4$}. Inset: Density correlations $C(k_y=0, x) = \langle \rho_{-k_y}(x) \rho_{k_y}(0)\rangle$ for  $V=1.1t$ and $V=2.0t$. Due to the finite cylinder circumference, translations are already weakly broken in the FQH phase. \textbf{(b)} The correlation length of the density mode, i.e. of operators with zero charge and spin $(Q, S_z)=(0, 0)$, peaks at a certain interaction strength $V$ which we identify as the critical point (indicated by a dashed line in all panels). By contrast, the spin correlation length $(Q, S_z)=(0, \hbar)$ remains small and does not grow when approaching the transition. \textbf{(c)} The ratio of $\xi_{2e}$ {[$(Q, S_z)=(2, 0)$] to $\xi_{1e}$ [$(Q, S_z)=(1, \hbar/2)$] } correlation length {is much larger than the trivially expected value of $0.5$ and peaks} close to the critical point, growing as a function of the bond dimension $\chi$.}
\label{fig:transition}
\end{center}
\end{figure*}

Having established and characterized the Halperin $(\bar 1\bar 12)$ and semion crystal states, we now study the topological quantum phase transition between them by increasing the nearest-neighbor interaction strength $V$, while keeping the on-site interaction $U=4t$ fixed.

First, we study the spontaneous breaking of discrete lattice translations as nearest-neighbor interactions are increased. 
We define the order parameter of the crystal as the static structure factor at the ordering wave vector $\mathbf{K}=(0, 4\pi/3a)^T$
\begin{equation}
    \Delta_\text{CDW} = \langle \rho(-\mathbf{K)} \rho(\mathbf{K})\rangle - \langle \rho(\mathbf{K})\rangle^2,
\end{equation}
with the density operator 
\begin{equation}
    \rho(\mathbf{k}) = \frac{1}{\sqrt{N}} \sum_{\mathbf{q}, \sigma}c^\dagger_{\sigma \mathbf{q}} c_{\sigma \mathbf{q}+\mathbf{k}}.
\end{equation}
We find the order parameter $\Delta_\text{CDW}$  to grow quickly near the critical point, consistent with a continuous transition; Fig.~\ref{fig:transition}(a). We remark that depending on the cylinder geometry, the FQH state may already weakly break translations due to a finite cylinder circumference. {This is, for instance, the case for the here considered $L_y=4$. Different geometries are discussed in Appendix~\ref{app:cylinders}}.

Next, we {study the spin and charge correlation lengths when tuning across the critical point. We find the} density correlation length, corresponding to the charge sector $(Q, S_z)=(0,0)$, to diverge at the critical point as a function of the bond dimension. In contrast, the spin correlation length, $(Q, S_z)=(0,\hbar)$, remains finite and quickly converges with bond dimension; see Fig.~\ref{fig:transition}(b). This suggests a finite spin gap across the transition, consistent with the continuous transition discussed in Sec.~\ref{sec:Transition}. 
Furthermore, the ratio between the charge $(Q, S_z)=(2, 0)$ and charge $(Q, S_z)=(1,\hbar/2)$ correlation length peaks at the critical point, growing with bond dimension and are, in particular, larger than the trivial expectation of 1/2, which arises when the charge-2 correlations would be only inherited from charge-1 correlations; Fig.~\ref{fig:transition}(c). This behavior is expected when charge-2 excitations become energetically favorable compared to spinful charge-1 excitations. 
{We emphasize that the large spin gap in both the FCI and the semion crystal protects these phases against perturbations. Accordingly, we expect them to be stable beyond the $\mathrm{SU(2)}$ symmetric model Eq.~\eqref{eq:microscopic-model} considered here. Since the spin gap remains finite across the transition, we anticipate this stability to be inherited upon doping in the vicinity of the critical point.} 

All together, these tensor network results provide strong evidence that our model realizes a continuous transition from a $(\bar{1}\bar{1}2)$ FQH state to a semion crystal, with the spin gap remaining open {and with enhancement of Cooper pair correlations across the transition. We leave the examination of the finite doping state to future numerical work.}

{We note that even in the strict limit of spinless fermions, the system could undergo the same FCI to semion crystal transition, enabling an identical anyon-superconductivity mechanism in its vicinity, just as in the spinful case. The relevant energy scale near the transition is set by a finite charge $Q=1e$ gap. Finding a microscopic model that stabilizes such spinless semion crystal and perhaps realizes a continuous transition from a 2/3 FCI phase, remains an exciting open challenge.}

\section{Experimental implications} \label{sec:ExperimentalImplications}

Let us now discuss the experimental outlook of our work. 

{\em Realizing the spinful Hubbard-Hofstadter Model with flat bands.} The microscopic model proposed Sec.~\ref{sec:Numerics} can be engineered in moir\'e materials by applying strong magnetic fields to a moir\'e lattice and utilizing layer quantum numbers, which are immune to the Zeeman splitting, as pseudospin~\cite{ZhangSU4_2021,Kuhlenkamp24}. Specifically, in {homo-bilayers of transition metal dichalcogenides (TMDs), twisted at a small angle after a 60 degree rotation (H-stacking) of one of the layers,} spin-valley locking suppresses tunneling between the layers.  
The electron therefore carries two independent quantum numbers: spin (which is locked to valley) and layer. A magnetic field then polarizes the electron spin~\cite{Kuhlenkamp24}, leaving a layer pseudospin that is immune to Zeeman splitting---providing exactly the ingredients of the two-component Hubbard-Hofstadter model~\cite{ZhangSU4_2021,Kuhlenkamp24}. Such models were found to realize the Integer Chern insulator to chiral spin liquid transition~\cite{Kuhlenkamp24,KuhlenkampThesis, divic_anyonsuperconductivity_2024}. Further,  realizing a robust 2/3 FQH state requires flat bands. In our model this is obtained by appropriate nearest-neighbor hoppings, a feature that will need to be tuned experimentally by knobs that control the electronic dispersion.

{\em Twisted MoTe$_2$.} The most striking experimental parallel is the observation of superconductivity in twisted MoTe${}_2$~\cite{xu_SC_2025}, in close proximity to the $2/3$ fractional quantum anomalous Hall insulator~\cite{cai_SignaturesFractional_2023,park_ObservationFractionally_2023}. {Hole doping the top valence band of twisted MoTe${}_2$} results in spontaneous valley polarization, and electrons occupy a single Chern band that hosts integer (filling $\nu=1$) and fractional ($\nu= 2/3, 3/5,\ldots  $) quantum anomalous Hall states. Between the $\nu=1$ and $\nu=2/3$ states, the system passes through an anomalous Hall metal and then a superconductor, while gate-tuned displacement fields stabilize additional valley-polarized insulators with zero Hall conductance. 

Therefore, from a coarse-grained point of view, both our setting and MoTe$_2$ host the fractional $2/3$ state, a proximate superconductor on doping, and nearby insulators without Hall conductance. However, some important differences between these systems should be noted, even at the level of symmetries. We have assumed spin rotation symmetry, while the valley polarized MoTe$_2$ has stongly-broken spin rotation symmetry due to spin valley locking. Can the two be related? The first observation is that since our model has a spin gap, weak breaking of spin-rotation symmetry leaves the phases intact. For instance, the topological order of the chiral spin liquid is stable when spin rotation symmetry is broken~\cite{Kuhlenkamp24}, despite the fact that spin transport is not quantized any longer. Next, we note that for the strong spin-rotation symmetry breaking relevant to tMoTe$_2$, our mechanism provides a general guideline: if the 2/3 FCI is in proximity to a specific charge density wave - the semion crystal - then the charge 2e/3 anyon becomes energetically cheaper than a pair of separated e/3 quasiparticles. Doping such an FCI should favor superconductivity. Repulsive interactions are indeed expected to drive FCIs towards CDW order, the key is to realize the correct type of CDW, (see Appendix~\ref{app:Jain_sequence}). A promising strategy is to search the phase diagram of MoTe$_2$ for insulating states that continue to break time reversal symmetry but exhibit a vanishing Hall response, which can be a potential candidate for this exotic CDW and examine if we have a continuous (or weakly first order) transition into the 2/3 FCI. Indeed, it has already been speculated that precisely such exotic CDWs could arise in MoTe$_2$~\cite{Song2024}. Our framework links that putative CDW directly to enhanced Cooper pairing in the neighboring FCI, providing a concrete route to accounting for superconductivity. 

As an alternative scenario, superconductivity in this system could instead evolve from the intervening chiral metal~\cite{xu_SC_2025, guerci2025_ChiralSC}. To distinguish these scenarios, it will be helpful to have more information on the energy ordering of quasiparticles. A simple (but indirect and imperfect) way to do this is via the observation (or absence) of daughter FCI plateaus, as the hierarchical sequence changes depending on which anyon is lowest energy. If the minimal charge 1/3 anyon is the cheapest, then one obtains the usual Haldane-Halperin hierarchy~\cite{HanssonHierarchyRevMod2017}, and several of these are observed for $\nu<2/3$. This is consistent with that being the lowest charge excitation on the electron doping (smaller $\nu$) side. However on the opposite side, incompressible states at fillings predicted by the 2/3 anyon hierarchy, such as $\nu=4/5$ would provide strong evidence for 2/3 anyons to be cheaper for hole doping (see Appendix~\ref{app:hierarchy}), regardless of the mechanism. This energy ordering would be consistent with the observation of superconductivity for $\nu>2/3$. 
{Other potential signatures of the energy ordering of anyons could come from spectroscopic probes~\cite{Inbar_2023, Peri_2024, Pichler2024, Hu2025, nuckollsSpectroscopyFractalHofstadter2025, Pichler2025}}{ or by measuring the electric charge via shot-noise measurements~\cite{dePicciotto1997,Saminadayar1997} or via single electron transitor measurements~\cite{Martin2004,Venkatachalam2011}}. 

In the MoTe$_2$ experiments, a re-entrant integer QAH plateau {(RIQAH)} was reported~\cite{xu_SC_2025} between the 2/3 and 3/5 FQAH states. Although additional confirmation is warranted—particularly since edge-equilibration effects could influence the measurement—taking the observation at face value, this state could potentially arise from the usual plateau transition driven by 1/3 anyons. In our formalism, {a different possibility is suggested.} Stacking a Chern-1 QAH layer on top of the 2/5 to semion-crystal transition also reproduces such a feature (see Appendix~\ref{app:RIAQH}). {The stacking converts the 2/5 into a 3/5 FQAH state, while the CDW, after stacking, reproduces the Hall conductance of the RIQAH. Note however, the presence of a background topological order in this scenario implies that this phase is more appropriately designated RIQAH$^*$. } 

{{\em Other candidate platforms:}  Multilayer rhombohedral graphene displays a zero-field FQAH series~\cite{LongJu2024,LongJuSc2025} and the microscopic Hamiltonian preserves full spin rotation symmetry.  Although superconductivity with broken time reversal symmetry has been observed in this system, it has so far appeared only in the moir\'eless regime where the fractional states have not yet been seen. Future theoretical studies retaining the spin degree of freedom will be needed to clarify their role, which may allow the mechanism outlined here to operate near the $\nu = 2/3$ state.  Experiments that search for the possible emergence of superconductivity near the 2/3 fractional state of multilayer rhombohedral graphene would be of great interest. Furthermore, high-field FCIs in graphene-hBN platforms~\cite{Spanton2018} provide another arena where continuous FCI–CDW transitions have been predicted~\cite{Lee2018,Gao2025}.

\section{Outlook \& Discussions} 
\label{sec:Outlook}
Superconductivity that emerges from doping conventional insulators requires that electrons first bind into Cooper pairs and those pairs then acquire long-range phase coherence.  In fractional Chern insulators (FCIs), the microscopic degrees of freedom are anyons rather than electrons. For anyons, the analog of the first step is to establish the right energy hierarchy for anyon charges. A finite density liquid of them then leads to phase coherence and superconductivity.   In this work, we have established how FCIs at the brink of a topological phase transition can supply the necessary energetic hierarchy that leads to anyon superconductivity. We have demonstrated that such transitions can be reached in microscopic electronic models with predominantly repulsive interactions.

In fact, the topological criticality mechanism previously discussed for integer Chern states leads precisely to binding of electrons, as long as one is on the Chern insulator side of the transition~\cite{divic_anyonsuperconductivity_2024}. Our fractional quantum Hall examples extend this mechanism to anyons of the FCI. Because the mechanism relies only on repulsive interactions and does not invoke time-reversal symmetry, it provides a realistic route to superconductivity in strongly correlated Chern systems.

One of the biggest surprises of our work is the numerical discovery of the semion crystal in simple Hubbard models at fractional filling. One may view this phase as a period-three electron CDW on which the residual spins form a chiral spin liquid, akin to the cluster-Mott physics proposed for 1T-TaS$_2$~\cite{LawLee2017}.  {Finding a stable semion crystal in such a minimal model highlights it as a viable state even competing with conventional crystals and FQH states.} Elucidating the properties and stability of such semion crystals at other {fillings or in the absence of a moir\'e potential (i.e. the formation of a Wigner crystal supporting a CSL) is an interesting direction.}%fractional fillings and lattice geometries is an interesting direction.

In future work, it will be essential to understand the effect of strong SU(2) symmetry breaking. In this context, an important step will be to study generalizations of our setting to higher Chern bands, which can smoothly connect to the SU(2) limit discussed in this work~\cite{RepellinPRR2020, LedwithPRL2022, DongPRR2023}.

Our analysis focused on Abelian quasiparticles, but FCIs supporting non-Abelian statistics exhibit even richer pairing possibilities~\cite{shi2025}. Determining how topological criticality operates in that context is an exciting theoretical challenge.

Recent calculations suggest that even in continuum Landau levels, the minimal-charge quasiparticles may form bound states~\cite{Gattu2025, Xu2025}. However, such binding is believed not to occur for 1/3 and 2/3 fillings~\cite{Jain2002} for Coulomb interactions. Investigating analogous binding phenomena in FCIs could further broaden the scope of anyon superconductivity. However, a recent numerical study of quasiparticle gaps in twisted MoTe$_2$ did not find evidence for binding of the e/3 quasiparticles~\cite{Regnault2025}. The energetic hierarchy of anyons in MoTe$_2$~\cite{park_ObservationFractionally_2023,park_ObservationFractionally_2023} or rhombohedral graphene~\cite{LongJu2024,LongJuSc2025}, and hence the question of relevance of anyon superconductivity remains an open question. A variety of experimental signatures could shed light on this problem from the existence of exotic hierarchy states (see Appendix~\ref{app:hierarchy}), thermal Hall responses, or spectroscopic probes of anyon gaps~\cite{Hu2025, Pichler2025, nuckollsSpectroscopyFractalHofstadter2025}. 

In summary, we have identified a new charge density wave, the semion crystal, that appears in a family of electronic models hosting FCIs.  We point out that a generic, interaction-driven route to anyon superconductivity exists, that leverages proximity of the FCI to the transition into the semion-crystal, which opens multiple avenues for advancing our understanding of unconventional superconductivity in topological flat bands.

\textit{\textbf{Note added.---}} While finalizing this manuscript, three interesting preprints on related topics appeared~\cite{Zhang2025, shi2025_2, Nosov2025}.

\textit{\textbf{Acknowledgments.---}} We thank J. Dong, P.J. Ledwith, J. Feldmeier, A. Imamoglu, G. Shavit, J. May-Mann, T. Helbig, and J. Alicea for insightful discussions. A.V. would like to thank Mike Zaletel, Stefan Divic, Zhaoyu Han, Darius Shi, Carolyn Zhang and Yahui Zhang for discussions and collaborations on related topics. C.K. acknowledges funding from the Swiss National Science Foundation (Postdoc.Mobility Grant No. 217884). A.V. is supported by a Simons Investigator award, the Simons Collaboration on Ultra-Quantum Matter, which is a grant from the Simons Foundation (651440, A.V.). 
 F.P. and M.K. acknowledge support from the Deutsche Forschungsgemeinschaft (DFG, German Research Foundation) under Germany’s Excellence Strategy–EXC–2111–390814868, TRR 360 – 492547816 and DFG grants No. KN1254/1-2, KN1254/2-1, the European Research Council (ERC) under the European Union’s Horizon
2020 research and innovation programme (grant agreement No 851161), the European Union (grant agreement No 101169765), as well as the Munich Quantum Valley, which is supported by the Bavarian state government with funds from the Hightech Agenda Bayern Plus.  

\textit{\textbf{Data availability.---}}
Data, data analysis, and simulation codes are available upon reasonable request on Zenodo~\cite{zenodo}.

\addtocontents{toc}{\string\tocdepth@munge}

\appendix
\section{Charge-2e superconductivity for generic singlet states}\label{app:generalCharge2SC}
Here we generalize the results of Sec.~\ref{sec:dopingVortices} of the main text to the fermionic $\nu=\frac{2}{2n\pm1}$ Halperin states in the vicinity of the transition to the semion crystal for generic $n$. The critical theory is again given by
\begin{equation}
\begin{aligned}
    \mathcal{L}_{\text{QCP}} &= \mathcal{L}_{\text{FQH}}[a+A] + \mathcal{L}_{
    \tilde{\varphi}}[\beta]-\frac{1}{2\pi}a\wedge\mathrm{\mathrm{d}\beta}\\   
    =-&\frac{n\pm1}{4\pi}(\alpha_\uparrow\wedge\mathrm{d}\alpha_\uparrow+ \alpha_\downarrow\wedge\mathrm{d}\alpha_\downarrow) - \frac{n}{2\pi}\alpha_\uparrow\wedge\mathrm{d}\alpha_\downarrow \\
    +& \frac{1}{2\pi}(a+A)\wedge\mathrm{d}(\alpha_\uparrow+\alpha_\downarrow) +  \mathcal{L}_{
    \tilde{\varphi}}[\beta]-\frac{1}{2\pi}a\wedge\mathrm{\mathrm{d}\beta}.
\end{aligned}
\end{equation}
As for $\nu=2/3$ and $\nu=2/5$, doped charges enter the system as vortices with charge $Q_{\tilde{\varphi}} = \frac{2}{2n\pm1}$. Thus we naturally find that upon doping $n_e = \frac{2}{2n\pm1} -\delta$ we introduced a finite density of vortices 
\begin{equation}
    n_{\tilde{\varphi}} = -\frac{2n\pm1}{2}\delta.
\end{equation}
The equations of motion determine the flux seen by the vortices to be:
\begin{equation}
    \frac{\mathrm{d}\beta}{2\pi} = \frac{\mathrm{d}\alpha_\uparrow + \mathrm{d}\alpha_\downarrow}{2\pi} = n_e = \frac{2}{2n\pm 1}-\delta.
\end{equation}
Due to the background density of electrons, the vortex field develops $2n\pm1$ valleys, while being at total filling  $\nu^\text{total}_{\text{vortices}} = n \pm 1/2$ with $n \in 2\mathbb{Z}$. 

At this filling, several possibilities to form quantum Hall states exist, which all give rise to superconductivity. One option is for the vortices to form a Laughlin state in each of the valleys, leading to the formation of $2n\pm1$ decoupled Laughlin states at level 2. This results in charge-2e superconductors which have a large residual topological order ($\det \tilde{K}_\text{doped} =16$ in addition to the TO of the superconductor when doping $\nu=2/5$). A perhaps simpler possibility is that vortices spontaneously form $n$ bosonic integer Quantum Hall states with the excess density being accommodated as a level $\pm2$ Laughlin state, with a block-diagonal K-matrix $K_\text{vortices}$. This choice generically breaks translations but leads to charge-2e superconductivity for all fillings without additional topological order, thereby naturally matching the superconductor we expect to form on top of the semion crystal. We suspect that in the limit of small electron densities; highly filled bosonic states of vortices become unstable to forming a superfluid and accommodate flux by forming a ``vortex lattice" themselves.

\begin{table}[]
    \centering
        \begin{tabular}{c|c|c|c}
        \toprule
        species & density &  flux  & filling \\
         & (w.r.t. lattice) & (per unit cell) & \\
        \hline
        $f_\sigma$ & $\frac{1}{2(n+m)}+\frac{\delta}{2}$  & $\pm(\frac{1}{2(m+n)} +\frac{\delta}{2})\Phi_0$ & $\pm2 \rightarrow \pm 2$ \\
        $g_1$ & $\frac{1}{m+n}+\delta$ & $(\frac{1}{m+n} \mp\delta)\Phi_0$ & $+1\rightarrow\mp 1$\\
        $g_2$ & $\frac{1}{m+n}+\delta$& $(\frac{1}{n+m} \pm\delta)\Phi_0$ & $+1\rightarrow\pm1$ \\
        $\vdots$ & $\vdots$ & $\vdots$ & $\vdots$ \\
        $g_{n-1}$ & $\frac{1}{m+n}+\delta$& $(\frac{1}{n+m} \mp\delta)\Phi_0$ & $+1\rightarrow\mp1$ \\
        $g_{n}$ & $\frac{1}{m+n}+\delta$& $(\frac{1}{n+m} \pm\frac{\delta}{2})\Phi_0$ & $+1\rightarrow\pm2$ \\
        \bottomrule
        \end{tabular}     
    \caption{Doping fermionic partons in the $(mmn)$ state for $m=n\pm1$. The flux is adjusted upon doping, such that the spinons remain in their original IQH state. The total Chern number (filling) before and after doping is shown.}
    \label{tab:doping_mmn_fermionic_general}
\end{table}

The same conclusions can again be drawn using fermionic partons only. {Generalizing Sec.~\ref{sec:dopingPartons} to general fermionic Halperin states}, the chargon splits up into $n$ fermionic partons
\begin{equation}
    b'(x) = \displaystyle\prod_{\alpha=1}^n g_\alpha,
\end{equation}
which introduces gauge fields $a_1, ..., a_{n}$
The flux of the spinons must change by 
\begin{equation}
   \frac{1}{2\pi} \nabla \times a_1 = \pm \frac{\delta}{2}
\label{eq:spinons_doped_flux_general}
\end{equation}
such that they remain in an IQH state. We show that choosing a change in flux for the gauge fields $a_\alpha$
\begin{equation}
   \frac{1}{2\pi} \nabla \times a_\alpha = \pm (-1)^{\alpha+1} \frac{\delta}{2}
\label{eq:general_doped_flux}
\end{equation}
results in a charge-2e superconductor. Note that the flux assignment Eq.~\eqref{eq:general_doped_flux} is consistent with no change in the external flux of $A$ and follows from demanding $\nabla \times (a_\alpha + a_{\alpha+1})=0$. Using the same arguments as discussed in the main text for $\nu=2/3$, we find that on both sides of the transition, the partons fill an integer number of Landau levels in the doped valley, as summarized in Table~\ref{tab:doping_mmn_fermionic_general}, where we assumed that translations are broken and doped partons enter in a unique valley. The resulting Lagrangian is
\begin{equation}
\begin{aligned}
        \mathcal{L} = \mp \frac{1}{4\pi}\Big[&\alpha_\uparrow \wedge \mathrm{d}\alpha_\uparrow
    + \alpha_\downarrow \wedge \mathrm{d}\alpha_\downarrow +\displaystyle\sum_{\alpha=1}^{n-1} (-1)^\alpha \gamma_\alpha \wedge \mathrm{d}\gamma_\alpha  \\
    & + \beta_1 \wedge \mathrm{d}\beta_1
    + \beta_2 \wedge \mathrm{d}\beta_2 \Big] + \mathcal{L}_\text{constr.}
\end{aligned}
\end{equation}
with
\begin{equation}
\begin{aligned}
        \mathcal{L}_\text{constr.} = 
    &\frac{1}{2\pi} \Big[ a_1 \wedge \mathrm{d}(\alpha_\uparrow + \alpha_\downarrow -\gamma_1) \\
    &+ \displaystyle\sum_{\alpha = 2 }^{n-1} a_\alpha \wedge \mathrm{d}(\gamma_{\alpha-1} - \gamma_\alpha) \\
    &  +  a_n \wedge \mathrm{d}(\gamma_{n-1} - \beta_1- \beta_2)  + (\beta_1 + \beta_2)\wedge \mathrm{d}A \Big].
\end{aligned}
\end{equation}

Integrating out the gauge fields $a_2, ..., a_{n-1}$ enforces the constraint $\gamma_1 = \gamma_2 = \cdots = \gamma_{n-1} \equiv \gamma$, while integrating out $a_1$ and $a_n$ leads to $\gamma = \beta_1 + \beta_2 = \alpha_\uparrow + \alpha_\downarrow$. After simplifying, we find a Lagrangian $\mathcal{L}^\text{doped}$ with $K$-matrix
\begin{equation}
   K^\mathrm{doped}=  \begin{pmatrix}
       \pm 1 & 0 & -1\\
        0 & \pm1 & -1\\
        -1 & -1& \pm2
    \end{pmatrix}, \; \mathbf{q} =\begin{pmatrix}
       1\\
        1 \\
        0 
    \end{pmatrix},
\end{equation}
describing the same charge-2e superconductor without leftover topological order as Eq.~\eqref{eq:doped_112_SC}. We remark that the partons $g_\alpha$ viewed alone, neglecting the coupling to the spinons, form a $\nu=\mp2$ bosonic IQH state.

\section{The hierarchy close to topological criticality} \label{app:hierarchy}
The energetic ordering of the anyons has important consequences not just for the formation of anyon superconductors, but also for the hierarchical quantum Hall states that can form at special fillings. To derive the hierarchy, it is commonly assumed that the lowest-energy quasi-particles carry the lowest possible charge. When the lowest lying anyon does not have the lowest charge, due to proximity to topological quantum criticality or otherwise, this strongly affects secondary Hall states that can form. We recall that the hierarchy of a quantum Hall state with K-matrix $K,\mathbf{q}$ is constructed by condensing quasi-particles with vector $\mathbf{l}$ in a quantum Hall state of level $2p$~\cite{WenClassification1992}. The resulting hierarchy state is described by $\tilde{K},\tilde{\mathbf{q}}$ with \begin{equation}
    \tilde{K} = \begin{pmatrix}
        K& -\mathbf{l} \\
        -\mathbf{l} & 2p
    \end{pmatrix}, \quad \tilde{\mathbf{q}} = (\mathbf{q},0).
\end{equation}

For the case of $\nu=2/3$ with $K_{2/3} = \begin{pmatrix}
        1& 2 \\
        2 & 1
    \end{pmatrix}, \mathbf{q}=(1,1)$, the usual hierarchy assumes that the charge $1/3$ anyons with $\mathbf{l} = (1,0)$ or $(0,1)$ enter the system. If instead the $2/3$ anyon with $\mathbf{l} =(1,1)$ is cheap we find hierarchical states at fillings
\begin{equation}
    \nu = \frac{2p}{ 3p - 1},
\end{equation}
which includes incompressible states with $\sigma_{xy}h/e^2$ of  $3/5,4/7,1/2,1,4/5,3/4$. We note that experimental observation of these daughter states could provide strong evidence for the $2/3$ anyon being energetically favorable. We also expect such hierarchy states to compete with anyon superconductivity at special fillings.

\section{Exotic mechanisms for a reentrant integer Quantum anomalous Hall state \label{app:RIAQH}} 

\begin{figure}
\begin{center}
\includegraphics[width=0.99\linewidth]{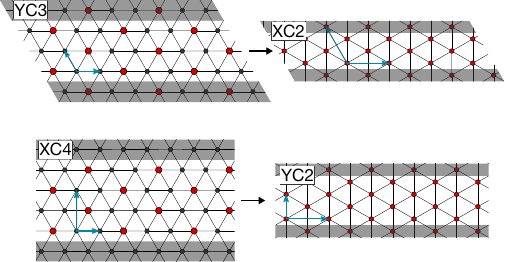}
\caption{\textbf{Different cylinder geometries.} The cylinders are infinite in one direction and have a finite circumference in the other. The translation vectors in the different geometries are shown in light blue. The red sites correspond to the charge order of the CDW and the effective cylinder geometry it forms. In the main text, we indicate the YC3 geometry with $L_y=3$ and the XC4 geometry with $L_y=4$.}
\label{fig:cylinders}
\end{center}
\end{figure}

\begin{figure*}
\begin{center}
\includegraphics[width=0.99\linewidth]{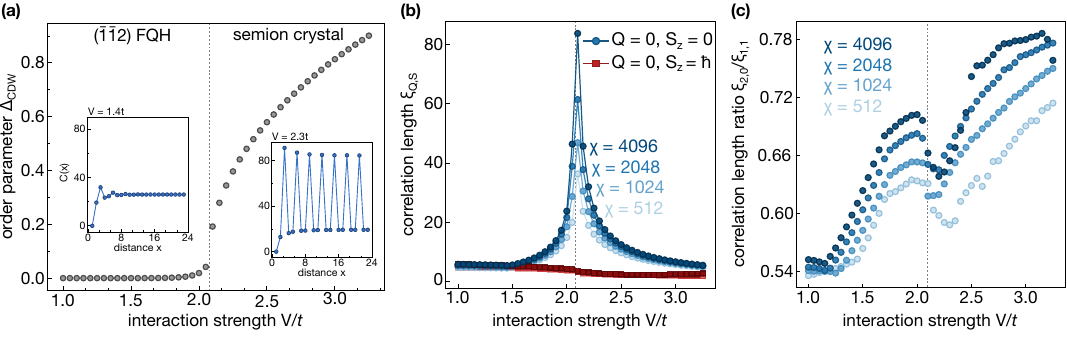}
\caption{
\textbf{Phase transition in the Hubbard-Hofstadter model for $L_y=3$.}  \textbf{(a)} Order parameter $\Delta_\text{CDW}$ as a function of the interaction strength $V$ for fixed $U=4t$, $\chi=2048$ and $L_y=3$. Inset: Density correlations $C(k_y=1, x) = \langle \rho_{-k_y}(x) \rho_{k_y}(0)\rangle$ for  $V=1.4t$ and $V=2.3t$. We show the $k_y=1$ correlations, since this corresponds to a cut through the Brillouin zone containing the ordering wave vector $\mathbf{K}$. For the $L_y=4$ geometry of the main text, the corresponding cut is $k_y=0$. In contrast to the FQH state on the $L_y=4$ cylinder discussed in the main text, translations are not even weakly broken in the $(\bar1 \bar12)$ phase on the $L_y=3$ cylinder. \textbf{(b)} The correlation length of the density mode, i.e. of operators with zero charge and spin $(Q, S_z)=(0, 0)$, peaks at a certain interaction strength $V$ which we identify as the critical point (indicated by a dashed line in all panels). The spin correlation length $(Q, S_z)=(0, \hbar)$ remains finite across the transition.  \textbf{(c)} The ratio of $\xi_{2e}$ [$(Q, S_z)=(2, 0)$] to $\xi_{1e}$ [$(Q, S_z)=(1, \hbar/2)$] correlation length is much larger than the trivially expected $1/2$ and peaks close to the critical point, growing as a function of the bond dimension $\chi$.}
\label{fig:transitionYC3}
\end{center}
\end{figure*}

Recent experiments have observed signatures consistent with a Reentrant Integer Quantum Anomalous Hall effect (RIQAH), appearing between $\nu=3/5$ and $\nu=2/3$. If one considers the $\nu=2/3$ state as a FQH state of holes forming out of a fully occupied band $\nu = 1-1/3$, it is natural to describe the transition $\nu=2/3 \leftrightarrow \text{RIQAH}$ as a crystallization transition $\nu=1-1/3 \leftrightarrow \nu=1+\text{CDW}$ for which the $Q=e/3$ anyon is active~\cite{Song2024}. An even more exotic option can arise by approaching the RIQAH state from $\nu=3/5$. This transition could take place if a $\nu=-2/5$ state transitions to a topological crystal as outlined in the main text, leaving behind an IQH response $\text{IQH}-2/5 \leftrightarrow \text{IQH+semion crystal}$. In such a case, the $Q=-2e/5$ anyon could be the lowest-lying excitation, thereby presenting a pathway to a superconducting state near the RIQAH phase.

\section{Chern numbers upon doping from Diophantine equation} \label{app:diophantine}
Given a triangular lattice model in some background flux $n_\Phi =\Phi/\Phi_0= p/q$ per unit cell, the Chern number in the $j-$th gap $C^j$ is given by the Diophantine equation~\cite{TKNN, Dana_1985, agazziColoredHofstadterButterfly2014}:
\begin{equation}
    C^j = (s j) \mod q \qquad \mathrm{with}\qquad (s p) \mod q = 1,
\end{equation}
where $s$ is the modular inverse of $p$. Or when expressed more concisely, for $n_e$ particles per unit cell, the total Chern number satisfies
\begin{equation}
    C n_\Phi - n_e = \mathbb{Z}.
\end{equation}
For the $(\bar{1}\bar{1}2)$ state discussed in the main text, the $g_2$-parton experiences a flux of 
\begin{equation}
   n_\Phi= \Phi_{g_2}/\Phi_0 = 1/3 - \delta/2
\end{equation}
and is at filling 
\begin{equation}
    n_{g_2} = n_e = 1/3 + \delta.
\end{equation}
We find that for this flux and filling, $C=-2$ solves the Diophantine equation for any doping $\delta$. We can also show it more explicitly if we write $\delta = 1/\ell$ with $\ell \in \mathbb{Z}$, which implies $\phi_{g_2} = \frac{2\ell -3}{6\ell}$ while having to fill $2\ell+6$ bands. The Chern number after accommodating all the $2\ell+6$ electrons is then:
\begin{equation}
\begin{aligned}
    C^{2\ell -6} &= [(2\ell - 3)^{-1} \cdot (6+2\ell)] \mod 6\ell \\
    &= -2 \mod 6\ell,
\end{aligned}
\end{equation}
where the first term on the RHS is understood as the modular inverse. The second line follows from $6l = 2 (2l-3) + (6+2l)$.
{We remark that simple solutions to the Diophantine equations resulting in superconductors are found only for specific choices of electron density $n_e$ and background-flux $n_\Phi$ per unit-cell. Consequently, we expect that the superconductor has to accommodate excess magnetic field by forming vortices. }

\begin{table*}
    \centering
    \begin{tabular}{lccc} % l = left, c = center, r = right
        \toprule
        mode & energy & momentum ($\times 2\pi/L_x$)& degeneracy \\ 
        \midrule
        ground state & $L_c=L_s=0$ & $0$ & $1\times (S_z=0)$ \\
        charge & $L_c=1$ & $-1$ & $1\times (S_z=0)$ \\
          & $L_c=2$ &$-2$ & $2\times (S_z=0)$ \\
          & $L_c=3$ &$-3$ & $3\times (S_z=0)$  \\
          & $L_c=4$ &$-4$ & $5\times (S_z=0)$  \\
        spin & $L_s=1$ & $2$ & $3 = 1 \times (S_z=0) + 1 \times (S_z=\pm1)$  \\
            & $L_s=2$ & $2$ & $4 = 2 \times (S_z=0) + 1 \times (S_z=\pm1)$ \\
          & $L_s=3$ & $3$ & $7 = 3 \times (S_z=0) + 2 \times (S_z=\pm1)$ \\  
          & $L_s=4$ & $4$ & $13 = 5 \times (S_z=0) + 3 \times (S_z=\pm1)  + 1 \times (S_z=\pm2)$ \\
        mixed & $L_c=L_s=1$ & $0$ & $3 = 1 \times (S_z=0) + 1 \times (S_z=\pm1)$ \\
            & $L_c=2, L_s=1$ & $-1$ & $6 = 2 \times (S_z=0) + 2 \times (S_z=\pm1)$ \\
            & $L_c=1, L_s=2$ & $1$ & $4 = 2 \times (S_z=0) + 1 \times (S_z=\pm1)$ \\
            & $L_c=L_s=2$ & $0$ & $6 = 4 \times (S_z=0) + 1 \times (S_z=\pm1)$ \\
        \bottomrule
    \end{tabular}
    \caption{Edge-mode counting for $Q=0$. The momentum is relative to the ground state. The total degeneracy of each mode is further resolved into the spin $S_z$ of the degenerate states.}
    \label{tab:countingQ=0}
\end{table*}

\begin{table*}
    \centering
    \begin{tabular}{lccc} % l = left, c = center, r = right
        \toprule
        mode & energy & momentum ($\times 2\pi/L_x$) & degeneracy \\ 
        \midrule
        ground state & $L_c=L_s=1/4$ & $0$ & $2=1\times (S_z=\pm1/2)$ \\
        charge & $L_c=1/4+1$ & $-1$ & $2=1\times (S_z=\pm1/2)$ \\
          & $L_c=1/4+2$ &$-2$ & $4=2\times (S_z=\pm1/2)$ \\
          & $L_c=1/4+3$ &$-3$ & $6=3\times (S_z=\pm1/2)$  \\
          & $L_c=1/4+4$ &$-4$ & $10=5\times (S_z=\pm1/2)$  \\
        spin & $L_s=1/4+1$ & $1$ & $2 = 1 \times (S_z=\pm1/2)$  \\
            & $L_s=1/4+2$ & $2$ & $6 = 2 \times (S_z=\pm1/2) + 1 \times (S_z=\pm3/2)$ \\
          & $L_s=1/4+3$ & $3$ & $8 = 3 \times (S_z=\pm1/2) + 1 \times (S_z=\pm3/2)$ \\  
          & $L_s=1/4+4$ & $4$ & $14 = 5 \times (S_z=\pm1/2) + 2 \times (S_z=\pm3/2) $ \\
        mixed & $L_c=L_s=1/4+1$ & $0$ & $2 = 1 \times (S_z=\pm1/2)$ \\
            & $L_c=1/4+2, L_s=1/4+1$ & $-1$ & $4 = 2 \times (S_z=\pm1/2) $ \\
            & $L_c=1/4+1, L_s=1/4+2$ & $1$ & $6 = 2 \times (S_z=\pm1/2) + 1 \times (S_z=\pm3/2)$ \\
            & $L_c=L_s=1/4+2$ & $0$ & $12 = 4 \times (S_z=\pm1/2) + 2 \times (S_z=\pm3/2)$ \\
        \bottomrule
    \end{tabular}
    \caption{Edge-mode counting for $Q=1$. The momentum is relative to the ground state. The total degeneracy of each mode is further resolved into the spin $S_z$ of the degenerate states. For the charge (spin) mode, $L_{s(c)}=1/4$ is fixed.}
    \label{tab:countingQ=1}
\end{table*}
\section{Counting for edge spectrum} \label{app:CountingEdgeSpectrum}

From the edge Lagrangian, Eq.~\eqref{eq:edgeLagrangian}, we obtain the edge Hamiltonian
\begin{equation}
    \mathcal{H}_\text{edge} = \int \frac{dy}{4\pi} [ v_c (\partial_y\varphi_c)^2 + v_s (\partial_y \varphi_s)^2], \label{eq:edgeHamiltonian}
\end{equation}
and the total momentum 
\begin{equation}
    \mathcal{P} = -\int\frac{dy}{4\pi} [(2n\pm1) (\partial_y \varphi_c)^2 \pm (\partial_y\varphi_s)^2]. \label{eq:edgeMomentum}
\end{equation}
We introduce a mode expansion for the bosonic edge fields~\cite{furukawa13}
\begin{equation}
\begin{aligned}
        \varphi_{c}(y) &= \frac{2\pi y}{L_y}\frac{Q}{\sqrt{4n \pm2}} \\ &\quad+ \displaystyle\sum_{\ell=1}^\infty\frac{1}{\sqrt{\ell(2n\pm1)}}(a_{c, \ell} e^{i k_\ell y} + a^\dagger_{c, \ell} e^{-i k_\ell y}) \\
        \varphi_{s}(x) &= \frac{2\pi y}{L_y}\sqrt{2} S_z + \displaystyle\sum_{\ell=1}^\infty\frac{1}{\sqrt{\ell}}(a_{s, \ell} e^{\pm i k_\ell y} + a^\dagger_{s, \ell} e^{\mp i k_\ell y})
\end{aligned}
\end{equation}
where $L_y$ is the length of the edge, $k_\ell = 2\pi \ell /L_y$, $Q$ is the total charge, $S_z$ the total spin, and $a_{c/s, \ell}$ are bosonic annihilation operators. 
Using this, Eqs.~\eqref{eq:edgeHamiltonian} and~\eqref{eq:edgeMomentum} can be written as
\begin{equation}
    \begin{aligned}
        &\mathcal{H}_\text{edge} = \frac{2\pi}{L_y} \bigg(\frac{v_c}{2n\pm 1} L_c + v_s L_s\bigg) \\
        &\mathcal{P} = -\frac{2\pi}{L_y}(L_c \pm L_s) \label{eq:edgeHamAndMomentum_Appendix}
    \end{aligned}
\end{equation}
with 
\begin{equation}
    L_c = \frac{Q^2}{4} + \displaystyle\sum_{\ell=1}^\infty \ell n_{c, \ell} \quad \text{and} \quad
    L_s = S_z^2 + \displaystyle\sum_{\ell=1}^\infty \ell n_{s, \ell}.
\end{equation}
Here $n_{c/s, \ell} = a^\dagger_{c/s, \ell}a_{c/s, \ell}$ is the occupation of the $\ell$-th charge/spin mode, $Q=N_\uparrow+N_\downarrow$ is the total charge, and $S_z = (N_\uparrow-N_\downarrow)/2$ the total spin. 

We outline how the edge spectrum and its degeneracies can be obtained for a fixed charge sector $Q$, focusing on $m=n-1$ relevant for the $(\bar{1}\bar{1}2)$ state. We first discuss $Q=0$, where there is a unique ground state with $L_c=L_s=0$ and momentum $\mathcal{P}=0$. There are three dispersing modes: A pure charge mode with $L_c\neq0$, $L_s=0$, a pure spin mode with $L_s\neq0$, $L_c=0$, and a mixed mode with $L_c\neq 0 \neq L_s$. For the charge mode, $L_s=0$ fixes $S_z=0$. The spectrum is given by
\begin{equation}
    \mathcal{H}_c = \frac{2\pi}{L_x} v_c L_c = \frac{2\pi v_c}{L_x}\displaystyle\sum_{\ell=1}^\infty \ell n_{c, \ell},
\end{equation}
resulting in the $(1,1,2,3,..)$ counting of a single bosonic chiral mode. The counting for the spin mode is more involved. We have 
\begin{equation}
        \mathcal{H}_s = \frac{2\pi}{L_x} v_s L_s = \frac{2\pi v_s}{L_x} \big[ S_z^2 + \displaystyle\sum_{\ell=1}^\infty \ell n_{s, \ell} \big].
\end{equation}
The lowest excitation corresponds to $L_s=1$ and has a threefold degeneracy: Either $S_z=0$ and $\sum_{\ell} \ell n_{s, \ell} = 1$ or $S_z = \pm 1$ and $\sum_{\ell} \ell n_{s, \ell} = 0$. For $L_s=2$ we find a fourfold degeneracy: two states from $S_z=\pm 1$ and $\sum_{\ell} \ell n_{s, \ell} = 1$ and two states from $S_z= 0$ and $\sum_{\ell} \ell n_{s, \ell} = 2$. The degeneracies for higher excited states can be worked out analogously and are summarized in Table~\ref{tab:countingQ=0}.
Finally, the lowest excited state of the mixed mode corresponds to $L_c=L_s=1$, with a threefold degeneracy: Either $S_z= 0$ and $\sum_{\ell} \ell n_{c, \ell} = \sum_{\ell} \ell n_{s, \ell} = 1$ or $S_z=\pm1$, $\sum_{\ell} \ell n_{s, \ell} = 0$, and $\sum_{\ell} \ell n_{c, \ell}=1$. Higher excited states are summarized in Table~\ref{tab:countingQ=0}. The counting in the $Q=1$ sector is given in Table~\ref{tab:countingQ=1}.

\section{Different cylinder geometries} \label{app:cylinders}

The transition to a semion crystal can only be observed on cylinder geometries, which are commensurate with the expected charge order of the crystal. We consider YC$-n$ and XC$-n$ cylinders, which have circumferences $L_{y} = n$.
For YC$-n$ cylinders, we require $n=3\ell$ to be an integer multiple of $3$ and for XC$-n$ cylinders $n=2\ell$ has to be an even integer, to accommodate the CDW with every third site occupied. The CDW on YC$-n$ cylinders forms an effective XC$-(n-1)$ geometry, while XC$-n$ reduces to YC$-(n-2)$; see Fig.~\ref{fig:cylinders}. For small cylinders, the details of the observed transition depend on the precise geometry, since we are effectively capturing a 1D transition. In the main text, we focused on the XC4 cylinder in Fig.~\ref{fig:transition}. The equivalent data for the YC3 geometry is shown in Fig.~\ref{fig:transitionYC3}.

{
\section{Computation of correlation lengths}

In our tensor network simulations, Sec.~\ref{sec:Numerics}, states are represented as infinite matrix product states (iMPS). In that case, the correlation function $C_A$ of any operator $A_c$ with charges $c$, can be expressed as~\cite{tenpy}
\begin{equation}
    C_{A_c}(x) = \displaystyle\sum_j a_j \lambda_{c, j}^x.
\end{equation}
where $a_j$ are some constant factors and $\lambda_{c, j}$ is the $j$-th eigenvalue of the transfer matrix in charge sector $c$. For large $x\rightarrow\infty$, one finds
\begin{equation}
    C_{A_c}(x) \propto e^{-x/\xi_c} \quad \text{with} \quad \xi_c = -\frac{1}{\log \lambda_{c, 1}}.
\end{equation}
Consequently, the correlations lengths $\xi_c$ are defined through the dominant eigenvalue of the transfer matrix in a given charge sector. In our iDMRG simulations, we have total charge $Q$, total spin $S_z$, and the momentum $k_y$ in $y$-direction, as conserved charges. The correlations length $\xi_{Q, S} = \max_{k_y} \xi_{Q, S, k_y}$ corresponds to the maximum (with respect to momentum $k_y$) correlation length for fixed charge $Q$ and spin $S_z$.
}

\section{Quantum critical point in terms of Cooper pairs}
\label{app:QCP_CooperPairs}
Here we extend the discussion of Sec.~\ref{sec:EffecTheoryCooperPairs} of the main text to the critical theory. At $\nu=2/3$ of the electrons, the critical theory can be mapped to a plateau transition between $\nu^{\text{Cooper}}=2/3$ and $\nu^{\text{Cooper}}=1/2$. This suggests that the transition can also be described by the following theory
\begin{equation}
\begin{aligned}    \mathcal{L}_\mathrm{FCI}^{\text{Cooper}} =& -\frac{2}{4\pi}\alpha \wedge\mathrm{d}\alpha + \frac{2}{2\pi}\alpha \wedge \mathrm{d}A +\frac{1}{2\pi}\alpha \wedge\mathrm{d}\tilde{\alpha}   \\
    - \frac{2}{4\pi}&\tilde{\alpha} \wedge \mathrm{d}\tilde{\alpha}
    + |D_{\beta}\varphi|^2 - V(\varphi)  -\frac{1}{2\pi}\tilde{\alpha}\wedge \mathrm{d}\beta,
\end{aligned}
\end{equation}
where $\alpha$ and $\tilde{\alpha}$ are emergent gauge fields describing the bosonic hierarchy state, and $\varphi$ is a scalar that drives the crystallization transition. The plateau transition between the $2/3$ hierarchy state of Cooper pairs (corresponding to the fermionic FCI) and the $1/2$ parent state (corresponding to the semion crystal) is now described by a condensation transition of the $\varphi$ boson. If $\varphi$ is condensed (Mott) we obtain the $\nu^{\text{Cooper}}=2/3$ ($\nu^{\text{Cooper}}=1/2$) state of Cooper pairs.
From the equations of motion we notice that the total electron density is given by $2\mathrm{d}\alpha/2\pi$, the density of $\varphi$ bosons is $\mathrm{d}\tilde{\alpha}/2\pi$ and the flux seen by the bosons $\mathrm{d}\beta = -3\mathrm{d}\alpha=-3 \mathrm{d}\tilde{\alpha}/2$. The superconductor is then obtained when the bosons form a FQH state corresponding to their filling of $-2/3$.

{
\section{Generalization to doping the Jain sequence}
\label{app:Jain_sequence}
While we focused on the spin-singlet Halperin sequence at filling $\nu= 2/(2n\pm 1), \; n$ an even integer, discussed in the main text, our theoretical analysis can also be generalized to Jain sates of spinless fermions. 

\textbf{\textit{Transitions of Jain states.---}}
Here we focus on the Jain sequence $\nu_\text{Jain} = \frac{p}{2p+1}, p\in\mathbb{Z}$, which has a simple parton description in terms of 
\begin{equation}
    e(x) = f_1(x) f_2(x)f_3(x),
    \label{eq:partons_jain}
\end{equation}
 where the partons are all at integer fillings $\nu_i = \lbrace p,1,1 \rbrace$~\cite{Sachdev_2023}. In contrast to the Halperin states, three fermionic partons suffice to describe the entire sequence. The Lagrangian of the Jain state after integrating out the constraints imposed by the parton construction can be expressed as
\begin{equation}
\begin{aligned}
    \mathcal{L}_\text{Jain} = -&\frac{2+\mathrm{sgn}(p)}{4\pi}\sum_{i=1}^{|p|}\alpha_i\wedge\mathrm{d}\alpha_i - \frac{2}{4\pi}\sum_{i\neq j}^{|p|}\alpha_i\wedge\mathrm{d}\alpha_j\\
    &+ \frac{1}{2\pi}A\wedge\mathrm{d}\sum_{i=1}^{|p|}\alpha_i.
\end{aligned}
\end{equation}
To favor superconducting doped states, we again construct possible continuous transitions of these states to CDW's by employing an extended parton construction
\begin{equation}
    e(x) = f_1f_2f_3b(x),
    \label{eq:extended_partons_jain}
\end{equation}
where $b$ is a hard-core boson as before. When the boson is condensed we recover $\mathcal{L}_{\text{Jain}}$, while a crystallized boson leads to an insulating CDW with topological order phase, which can be thought of as an anyon crystal (AX) with action
\begin{equation}
    \mathcal{L}_{\text{AX}} = -\frac{2\mathrm{sgn}(p)}{4\pi} \sum_{i=1}^{|p|-1}\alpha_i\wedge \mathrm{d}\alpha_i -\frac{\mathrm{sgn}(p)}{4\pi}\sum_{i\neq j}^{|p|-1} \alpha_i \wedge \mathrm{d}\alpha_j,
\end{equation}
with $|\mathrm{det}K_{\text{AX}}| = |p|$, leading to a residual topological order of the crystal. Note that the vortex which becomes soft has charge $Q_\Phi = e\nu = e\frac{p}{2p+1}$, which suggests a path to anyon-SC when $p\in 2\mathbb{Z}$. Perhaps even more exotic, when $p \in 2\mathbb{Z}+1$ the system could naturally form a charge-$p$ Fermi liquid. 

\textbf{\textit{Relation to plateau transition of composite objects.---}}
As for the singlet states, stacking with \textit{integer} phases highlights a dual description in terms of pleateau transitions of composite objects. Imagine stacking the critical theory with a $\nu = -p$ Chern insulator. The resulting insulator insulator transition involves a change of Hall conductivity 
\begin{align}
    \sigma_{xy} &: \quad \frac{p}{2p+1} \longleftrightarrow  0 \\
    \sigma^{\text{stacked}}_{xy}  &: \quad \frac{-2p^2}{2p+1} \longleftrightarrow  -p.
\end{align}
Such a transition can be interpreted as a natural transition between a $\nu_\text{composite} = -1/p$ Laughlin state of charge-$p\cdot e$ particles and its hierarchy state at filling $\nu_\text{composite}=-2/(2p+1)$ describing the AX and the Jain state (the charge vector is $p$). As such the QCP is equivalent to the plateau transition between
\begin{equation}
\nu_{\text{composite}} : \quad \frac{-2}{2p+1} \longleftrightarrow  -\frac{1}{p} 
\end{equation}
Note that this works out for both even and odd $p$, as the AX phase can be reinterpreted as a bosonic or fermionic Laughlin state, respectively. Similar transitions naturally appear for electrons with $p$ flavors. It can be checked that the topological order of the $\nu=\frac{p}{2p+1}$ is consistent with a fermionic SU($p$) singlet state, for which a corresponding SU$(p)_1$ 'spin' liquid state exists as well.

Based on the arguments presented in the previous sections we expect transitions of Jain states involving even $p$ to yield a charge $|p|$ superconductor. %while odd $p$ Jain states are likely to result in charge $|p|$ Fermi liquid states upon doping.

\textbf{\textit{Doping the Jain state.---}}
For $p=\pm2$ the physics is identical to the one discussed for the Halperin singlets, albeit in the absence of $\mathrm{SU(2)}$ symmetry. Nevertheless the resulting superconductor is found to have chiral central charge $c_-=\pm 2$.

The next interesting superconductor appears at $\nu= 4/7$ ($p=-4$). Repeating a derivation similar to the one in the main text, we find that the vortices are at filling $\nu_\Phi = 7/4$ and have $7$ flavors. When translational invariance is spontaneously broken upon condensation, one of the vortices can naturally form a hierarchy state at $ \nu=2-\frac{1}{4}$ around the bosonic integer quantum Hall state  with $K$-matrix 
\begin{equation}
     K =  \begin{pmatrix}
        0 & 1 & 0\\
        1 & 0  & 0 \\
        0 & 0 & -4
    \end{pmatrix}, \qquad \mathbf{q} = \begin{pmatrix}
        1 \\
        1  \\
        1 
    \end{pmatrix}.
\end{equation}
\\

Upon combining this with the fermionic sector, we indeed find that the resulting topological superconductor is charge-$4e$ with chiral central charge $c_- = -4$ and no residual topological order.

As in the previous cases, any other gapped vortex state at this filling would also yield a superconductor, although it could generically have topological order and in that case would have to be classified as an SC$^*$ phase.
}
\bibliography{library}

\end{document}